\documentclass[a4paper,fleqn,usenatbib]{mnras}

\usepackage{newtxtext,newtxmath}

\usepackage[T1]{fontenc}

\DeclareRobustCommand{\VAN}[3]{#2}
\let\VANthebibliography\thebibliography
\def\thebibliography{\DeclareRobustCommand{\VAN}[3]{##3}\VANthebibliography}

\usepackage{graphicx}	
\usepackage{bm}
\usepackage{amsmath}	

\title[\textit{HST} Observations of {[OIII]} Emission in Nearby QSO2s]{Hubble Space Telescope Observations of [O~III] Emission in Nearby QSO2s: Physical Properties of the Ionised Outflows}

\author[Trindade Falcão et al.]{
Anna Trindade Falcão ,$^{1}$\thanks{E-mail: anna.trindade04@gmail.com}
S. B. Kraemer,$^{1}$
T. C. Fischer,$^{2}$
D. M. Crenshaw,$^{3}$
M. Revalski,$^{2}$
H. R. Schmitt,$^{4}$
\newauthor
M. Vestergaard,$^{5,6}$
M. Elvis,$^{7}$
C.M. Gaskell,$^{8}$
F. Hamann,$^{9}$
L. C. Ho,$^{10}$
J. Hutchings,$^{11}$
R. Mushotzky,$^{12}$
H. Netzer,$^{13}$
\newauthor
T. Storchi-Bergmann,$^{14}$
T.J. Turner,$^{15}$
M.J. Ward$^{16}$
\\
$^{1}$Institute for Astrophysics and Computational Sciences, Department of Physics, The Catholic University of America, Washington, DC 20064, USA\\
$^{2}$Space Telescope Science Institute, Baltimore, MD 21218, USA\\
$^{3}$Department of Physics and Astronomy, Georgia State University, Astronomy Offices, 25 Park Place, Suite 600, Atlanta, GA 30303, USA\\
$^{4}$Naval Research Laboratory, Washington, DC 20375, USA\\
$^{5}$Dark Cosmology Centre, Niels Bohr Institute, University of Copenhagen, Jagtvej 128, 2200 Copenhagen N, Denmark\\
$^{6}$Steward Observatory and Department of Astronomy, University of Arizona, 933 N. Cherry Avenue, Tucson AZ 85721\\
$^{7}$Harvard-Smithsonian Center for Astrophysics, 60 Garden St., Cambridge, MA 02138, USA\\
$^{8}$Department of Astronomy and Astrophysics, University of California, Santa Cruz, CA 95064, USA\\
$^{9}$Department of Physics and Astronomy, University of California, Riverside, CA 92507, USA\\
$^{10}$Kavli Institute for Astronomy and Astrophysics, Peking University; School of Physics, Department of Astronomy, Peking University; Beijing 100871, China\\
$^{11}$Dominion Astrophysical Observatory, NRC Herzberg Institute of Astrophysics, 5071 West Saanich Road, Victoria, BC, V9E 2E7, Canada\\
$^{12}$Department of Astronomy, University of Maryland, College Park, MD 20742, USA\\
$^{13}$School of Physics and Astronomy, Tel Aviv University, Tel Aviv 69978, Israel\\
$^{14}$Departamento de Astronomia, Universidade Federal do Rio Grande do Sul, IF, CP 15051, 91501-970 Porto Alegre, RS, Brazil\\
$^{15}$Department of Physics, University of Maryland Baltimore County, 1000 Hilltop Circle, Baltimore, MD 21250, USA\\
$^{16}$Centre for Extragalactic Astronomy, Department of Physics, University of Durham, South Road, Durham DH1 3LE, UK\\
}

\date{Accepted XXX. Received YYY; in original form ZZZ}

\pubyear{2020}

\begin{document}
\label{firstpage}
\pagerange{\pageref{firstpage}--\pageref{lastpage}}
\maketitle

\begin{abstract}
We use Hubble Space Telescope  (\textit{HST})/ Space Telescope Imaging Spectrograph (STIS) long-slit G430M and G750M spectra to analyse the extended [O~III] $\lambda5007$ emission in a sample of twelve nearby ($z$ $<$ 0.12) luminous ($L_{bol} > 1.6 \times 10^{45}$ ${\rm erg}$ ${\rm s^{-1}}$) QSO2s. The purpose of the study is to determine the properties of the mass outflows of ionised gas and their role in AGN feedback. We measure fluxes and velocities as functions of radial distances. Using Cloudy models and ionising luminosities derived from [O~III] $\lambda5007$, we are able to estimate the densities for the emission-line gas. From these results, we derive masses of [O~III]-emitting gas, mass outflow rates, kinetic energies, kinetic luminosities, momenta and momentum flow rates as a function of radial distance for each of the targets. For the sample, masses are several times $10^{3}$~${\rm M\textsubscript{\(\odot\)}}$ - $10^{7}$~${\rm M\textsubscript{\(\odot\)}}$ and peak outflow rates are $9.3\times 10^{-3}$ ${\rm M\textsubscript{\(\odot\)}}$ ${\rm yr^{-1}}$ to $10.3$ ${\rm M\textsubscript{\(\odot\)}}$ ${\rm yr^{-1}}$. The peak kinetic luminosities are $3.4\times 10^{-8}$ to $4.9\times 10^{-4}$ of the bolometric luminosity, which does not approach the $5.0\times 10^{-3}$ - $5.0\times 10^{-2}$ range required by some models for efficient feedback. For Mrk 34, which has the largest kinetic luminosity of our sample, in order to produce efficient feedback there would have to be 10 times more [O~III]-emitting gas than we detected at its position of maximum kinetic luminosity. Three targets show extended [O~III] emission, but compact outflow regions. This may be due to different mass profiles or different evolutionary histories.

\end{abstract}




\section{Introduction}

An Active Galactic Nucleus (AGN) is a compact region at the centre of a galaxy that emits a significant amount of energy over much of the electromagnetic spectrum, and whose spectral characteristics indicate that the energy source is non-stellar. Such objects have been observed in the infrared, X-ray, radio, microwave, gamma-ray and optical/ultraviolet wavebands. \par 
Accreting supermassive black holes (SMBHs) are believed to be the central engines that power all AGN. This accretion is the result of mass inflows to the central SMBH that can be triggered both from outside the galaxy, via interactions with companions, or from inside it, via secular processes \citep{SS2019}. The generally adopted picture is that of a small continuum source, associated with the mass accretion flow that feeds the SMBH, surrounded by a much larger emission-line region \citep{OF,C2010,K2012}. The extended (10s - 1000s of pc) ionised gas in AGN is referred to as the "Narrow Line Region"(NLR). Here the typical electron density is $10^{2}$-$10^{6}$ ${\rm cm^{-3}}$ \citep{PT}, and the gas velocity is 300-1100 km ${\rm s^{-1}}$. The radiation released by the accretion flow to the SMBH can interact with the interstellar stellar medium of the host galaxy, ionising and accelerating the gas. This process may regulate the SMBH accretion rate. The relationship between the SMBH mass and the stellar velocity dispersion of its galaxy bulge (\citealt{KH2013}, and references therein) is credited to the action of the AGN quenching star formation and evacuating gas from the bulge, a process referred to as "AGN feedback" \citep{B2004}.\par

Various physical scenarios for effective feedback have been suggested. These include quenching star formation through negative feedback \citep{WZ2016}, triggering star formation through positive feedback \citep{S2013,M2017} or more complex interactions \citep{ZB2017}.\par

AGN feedback certainly exists in radio loud AGNs, whose powerful jets are highly collimated, but they occur in only 5-10\% of the AGN population \citep{Ra2009}. Meanwhile, winds are prevalent in most AGN \citep{M2013,G2014,W2016}. AGN winds are frequently observed as UV and X-ray absorption lines blueshifted with respect to their host galaxies, originating in gas within tens to hundreds of parsecs from the central SMBH \citep{C2003, V2005, CK2012, KP2015}, or emission-line gas in AGN narrow-line regions \citep{CK2005, C2010, MS2011, F2013, F2014, BW2016, N2016}. Recent studies \citep[e.g.,][hereafter\defcitealias{F2018}{F2018}\citetalias{F2018}]{F2018} question how effective AGN feedback is on galactic-bulge scales, as required in a star-formation quenching, negative feedback scenario. Therefore it is important to quantify its impact, which can be accomplished by characterising the physical properties of these outflows, such as mass, velocity, mass outflow rate, and kinetic energy. \par

While previous studies suggested that the power of the outflows scale with luminosity \citep{GB2008}, some ground-based studies of QSO2s \citep{G2011, L2013,H2014,ME2015} have found that powerful and very extended outflows detected by the optical [O~III] $\lambda$5007 emission line are extremely rare. This raises the question of whether kpc-scale AGN winds exist in most QSO2s. The answer to this question can be decisive on the matter of whether outflows are a critical component of quasars feedback and hence the evolution of galaxy bulges, or if the star formation is quenched in bulges by other means.\par

\defcitealias{F2018}{F2018}\citetalias{F2018} obtained \textit{HST} imaging and spectroscopy of 12 of the 15 most luminous targets at z $\leq$ 0.12 from the \citet{R2008} sample of QSO2s, through Hubble Program ID 13728 (PI: Kraemer) and archival observations of Mrk 34 (see Table \ref{tab:Table_1}). They measured [O~III] velocities and line profile widths as a function of radial distance in order to characterise mass outflows in these QSO2s. \par 

In regard to the morphology of the sample, \citetalias{F2018} found that in some of the targets the [O~III] region is very extended, such as in FIRST J120041.4+314745, which has a maximum radial extent in its [O~III] image, $R_{max}$, of 5.92 kpc. Meanwhile, other targets present a very compact morphology, such as 2MASX J14054117+4026326 which possesses a  $R_{max}$ = 0.88 kpc. \par 

Regarding the kinematics of the ionised gas, \citetalias{F2018} showed that the extent of the outflows, $R_{out}$, in most of the sample, is relatively small compared to the overall extent of the [O~III] emission region, $R_{max}$, with an average $R_{out}$/$R_{max}$ = 0.22, except for Mrk 34, for which $R_{out}$/$R_{max}$ $\sim$ 1 (see Table \ref{tab:Table_1}). They found that one can categorize the influence of the central AGNs in different regions, as a function of distance from the nucleus. In the inner region, the emission lines have multiple components and include velocity profiles that differ from rotation, i.e., with high central velocities and high {\textit {Full Width at Half Maximum}} (FWHM), and hence are consistent with outflows. At greater distances, gas is still being ionised by the AGN radiation but emission lines exhibit low central velocities with low FWHM, consistent with rotation of the host galaxy. In addition, \citetalias{F2018} identified a third kinematic component, namely, gas with low central velocities, but high FWHM. They refer to this as "disturbed" kinematics and suggested that AGN activity may be disrupting gas without resulting in radial acceleration.\par

Previous work \citep{F2017, WM2018} explored the idea that the outflows are radiatively accelerated, although we see in our targets a rotational component as well. At large distances, this is consistent with the fact that the flux of radiation is low, but the gravitational deceleration due to the enclosed stellar mass is large, therefore the gas cannot be radially accelerated. However, at smaller distances, the rotation component may simply be gas that has not been exposed to the AGN radiation long enough to be accelerated.\par 

In this study, we use the same data as \citetalias{F2018}, and we extend their analysis by computing masses, mass outflow rates, kinetic energies, kinetic energy rates, momenta and momentum flow rates for the same sample of QSO2s. Throughout this paper we adopt a flat $\Lambda$CDM cosmology
with $H_{0} =71{\rm km~s^{-1}}$ ${\rm Mpc^{-1}}$, $\Omega_{0}$ = 0.28 and $\Omega_{\lambda}$= 0.72.

\section{Sample, Observations and Measurements}

\subsection{\textit{HST} Observations}
\label{sec:sec_2.1} 
We use medium dispersion spectra\footnote[1]{R=$\lambda/\Delta \lambda ~ \sim$ 5000–10,000 \citep{WG1998}} to characterise the physical properties and kinematics of the emission-line gas, along with [O~III] imaging to determine the ionised gas mass. As already discussed in \citetalias{F2018}, the observing program for our sample was performed in a two-step process: first, we obtained narrow-band images of each AGN to determine ideal STIS position angles and, later, a spectroscopic observation. To obtain the images for our sample, FR505N or FR551N filters were used, chosen depending on the redshift of each target to observe [O~III], with the Wide-Field Channel (WFC) of \textit{HST}/Advanced Camera for Surveys (ACS). The FR647M filter was used to obtain the continuum observations.\par	

The long slit spectra used in this study were obtained with STIS using either the G430M or G750M gratings to study the [O III] kinematics, employing a $52''\times 0.2''$ slit oriented along the major axis of the NLR.  

\subsection{Spectral Fitting}
\label{sec:sec_2.2} 
\begin{figure}
  \centering
  \includegraphics[width=0.5\textwidth]{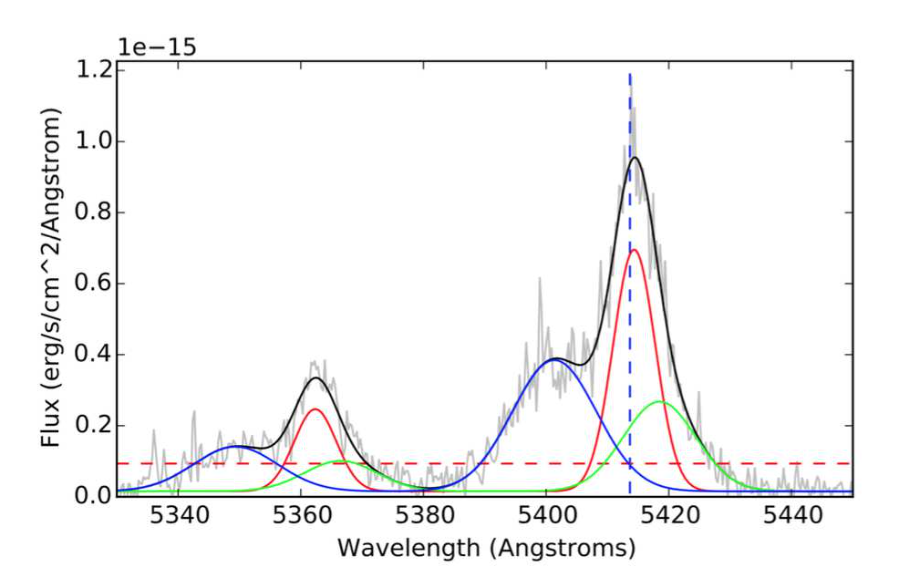}
\caption{\label{fig:Fig_1}[O~III] $\lambda4959$,$\lambda5007$ emission-line component fitting example over the continuum peak in 2MASX J08025293+2552551. The continuum peak refers to the brightest row in the 2D CCD data, where, if one takes a vertical, 1D cut along the image (avoiding the emission lines), that row would be the peak of the flux distribution. [O~III] $\lambda4959$ fit parameters are fixed to be identical to [O~III]
$\lambda5007$ fit parameters, with line flux fixed to be 1/3 of [O III] $\lambda5007$ flux. Grey line represents STIS spectral data. Solid black line represents the total model. Red, blue, and cyan lines represent individual Gaussians. Vertical dashed blue line represents the [O~III] $\lambda5007$ wavelength at systemic velocity. Horizontal dashed red line represents the 3$\sigma$ continuum-flux lower limit for Gaussians in our fitting. Figure from Fischer et al. 2018.}
\end{figure}

We fit the emission line in each row (i.e., in the spatial direction) of the STIS spectral image with Gaussians in order to obtain the [O~III] velocities, relative to systemic, and fluxes. We employ a Bayesian fitting routine, discussed by \citet{F2017}, that uses the Importance Nested Sampling algorithm in the MultiNest library \citep{FH2008, Fr2009, Fr2013, B2014} to compute the logarithm of the evidence, $\ln(Z)$, for each model, as shown in Figure \ref{fig:Fig_1}.\par 
The models are run for zero Gaussian components, i.e., no [O~III] emission, and then for one Gaussian component. If the one-component model is chosen over the zero component model\footnote[2]{The one component model has to have a significantly better evidence value, $|\ln\frac{Z_1}{Z_0}>5|$, in order to be chosen over the zero component model.} the data are analysed with a two-component model, and the process is repeated until the more complex model $(\ln Z_{n+1})$ is no longer chosen over the previous one $(\ln Z_n)$. The uncertainty in flux for each line is calculated from the residuals between the data and the fit. However, the flux uncertainties are small compared to those discussed in section \ref{sec:sec_2.4}.

\subsection{[O~III] Image Analysis}
\label{sec:sec_2.3} 

In order to account for the mass of [O~III] emitting gas outside of the area sampled by the STIS slit, we use a continuum-subtracted [O~III] emission line image of the entire NLR for each target. The flux calibration included scaling by the filter bandpass. These correction factors are equal to 2\% of the linear ramp filter wavelength, or approximately 100 \AA, for each image \citep{Ry2019}.\par

Our measurements and velocities discussed in section  \ref{sec:sec_2.2} are deprojected according to the analysis of the Sloan Digital Sky Survey (SDSS) images, as described in \citetalias{F2018}. An example of the [O~III] image and azimuthally summed flux profile are shown in Figure 2, for FIRST J120041.4+314745. \par
\begin{figure*}
  \centering
 \begin{minipage}[b]{0.4\textwidth}
  \includegraphics[clip,width=7.55cm, height=5cm]{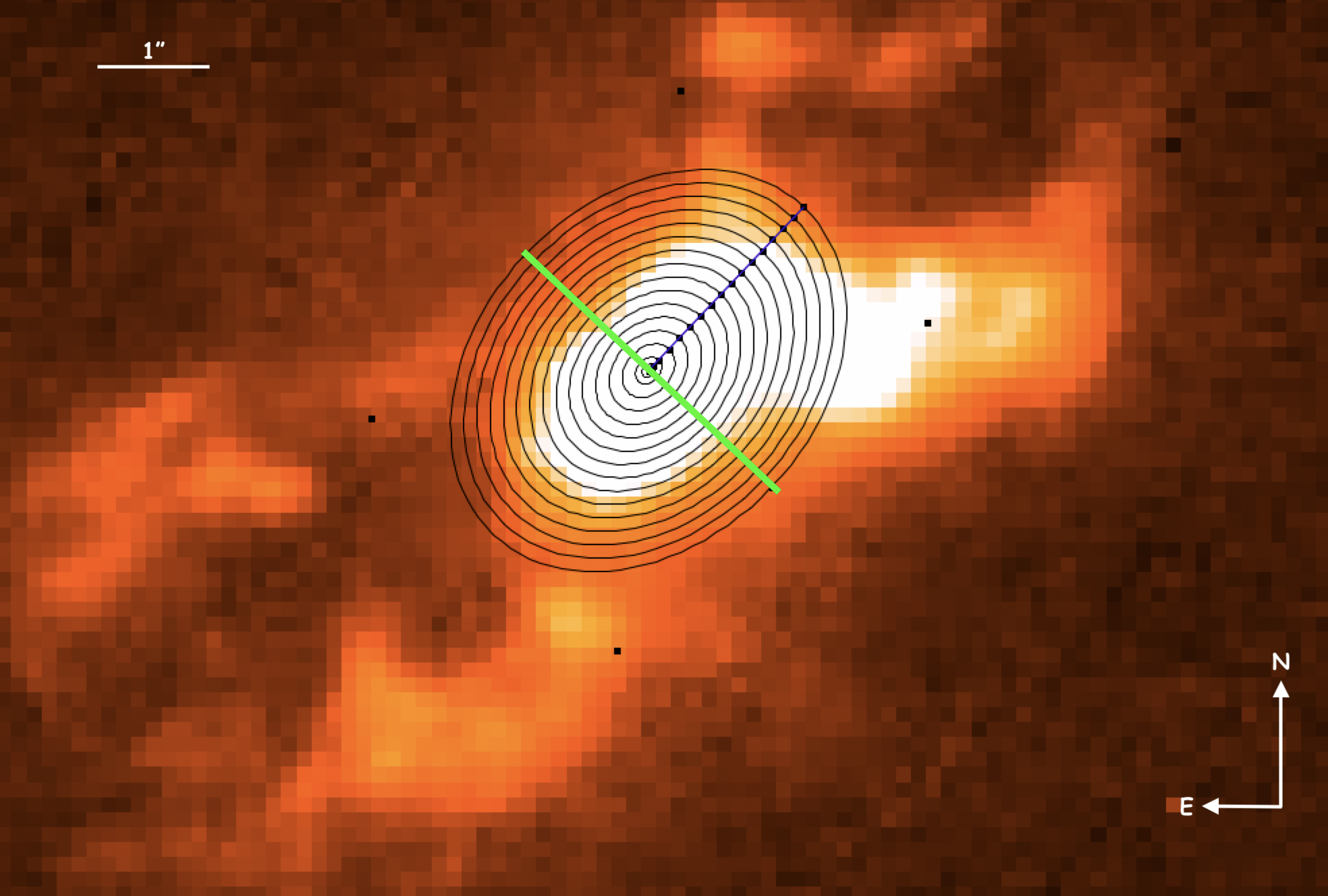}
 \end{minipage}\qquad
 \begin{minipage}[b]{0.55\textwidth}
  \includegraphics[clip,width=8.5cm,height=5.35cm]{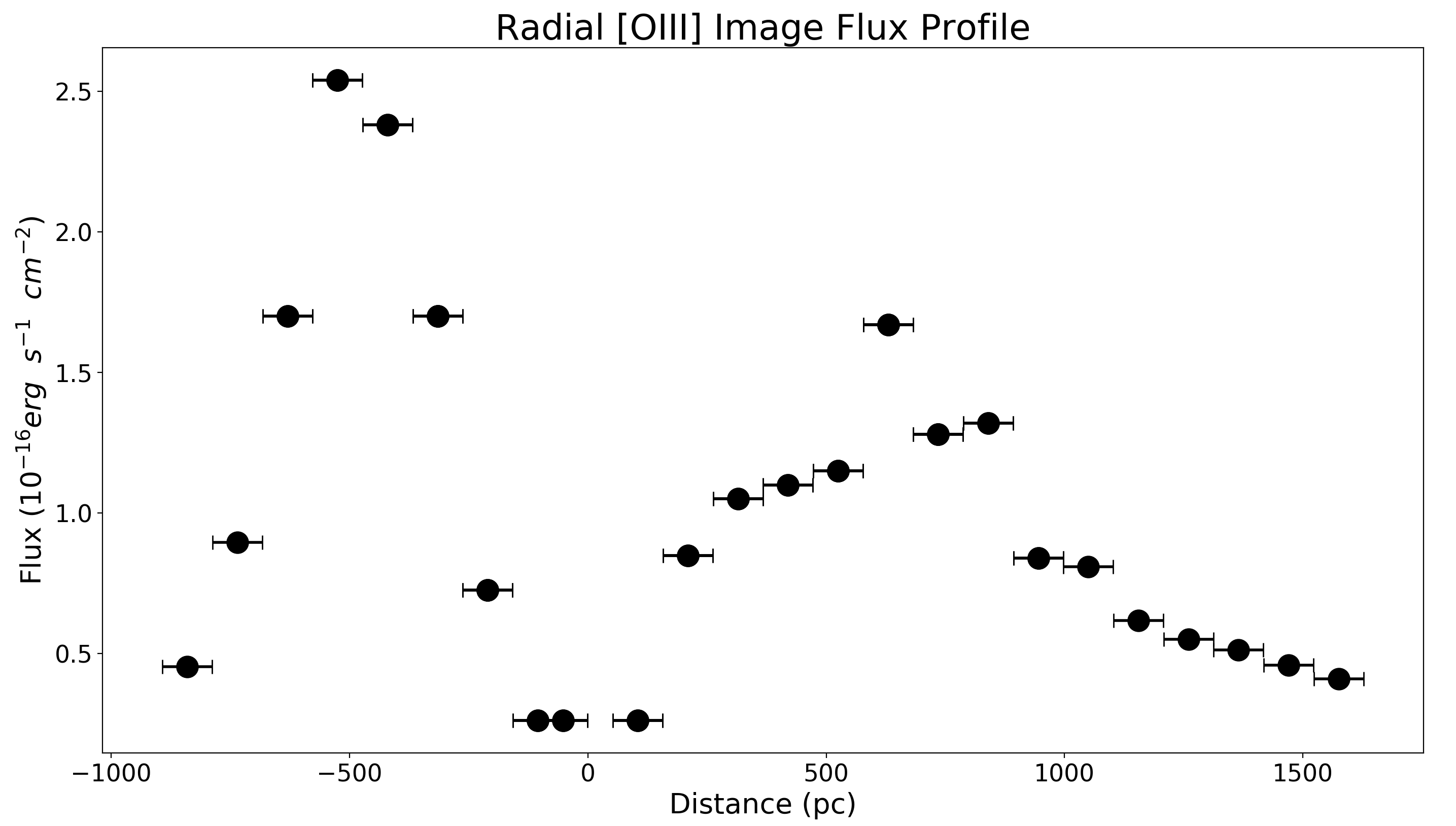}
 \end{minipage}\qquad 
\caption{The left panel shows the \textit{HST} [O~III] image of FIRST J120041.4+314745 with superimposed elliptical semi-annuli representing rings of constant distance from the nucleus. The green line represents the positions where we divide the two semi-ellipses. The orthogonal black line represents the direction of the slit. The right panel is the [O~III] annuli fluxes oriented along the major axis of the ellipse. The negative values are to the southeast and the positive values are to the northwest. Ellipses very close to the nucleus have a very small size, leading to low fluxes in the centre positions. Fluxes at positions $>$ 900pc (to the southeast side) are too low to be detected. The radial axes in the right panel are asymmetric because the kinematic fits were limited to this range.}
\label{fig:Fig_2}
\end{figure*}

 
To determine the total [O~III] flux as a function of distance from the nucleus, we use the Elliptical Panda routine, in the SAOImage DS9 Software \citep{JM2003}. Following \citetalias{F2018} (see sections 2.5 and 3.1 of \citetalias{F2018}), we assume that the host galaxies are disc galaxies and that the inclination of their discs to our line of sight can be obtained from the ellipticities of their isophotes. The ellipticities and position angles (PAs) of our targets are given
in Table 2 of \citetalias{F2018} along with the PAs of the ionization comes. Following \citet{F2017} and \citetalias{F2018}, we furthermore assume that most of the AGN-ionized gas structure in type-2 AGNs
lies in their host disks, and that we can thus use the host disc orientation assumed for each galaxy to deproject \textit{HST} measurements for our QSO2 sample to determine true physical distances in the plane of the host galaxy. Annuli of constant distance from the centres of the galaxies are ellipses with their major axes in the PA as given in Table 2 of \citetalias{F2018}. We add up the [O~III] flux in a series of elliptical annuli of radius $\delta r$ (see Table \ref{tab:Table_1}), where $\delta r$ is the deprojected length along the slit for each extraction. These annuli are illustrated for one of our targets in the left-hand panel of Figure \ref{fig:Fig_2}. We divide each elliptical annulus in two with one half for each side of the ionization cone.\par 
  
\subsection{Constraints on Luminosity}
\label{sec:sec_2.4} 

In Type 2 AGNs, the inner region, nearest to the SMBH, is hidden from view \citep{A1993}, therefore one has to use indirect indicators to estimate the bolometric luminosity, except in cases where the X-ray absorber is not Compton-thick. One method is to use the total [O~III] luminosity, $L_{[O~III]}$, and a bolometric correction factor \citep[e.g.,][]{Hc2004}, but there is some uncertainty in the value of the factor depending on the extinction \citep{L2009} and the Eddington ratio \citep{Du2020}. \par 

In order to estimate the extinction towards the NLRs of the QSO2s, we retrieved the fluxes of $H\alpha$ and $H\beta$ from SDSS \citep{Ah2019}. The observed ratios of $H\alpha$/$H\beta$ range from  4.7, for 2MASX J07594101+5050245, to  3.5, for FIRST J120041.4+314745. In AGN, the intrinsic ratio can range from the theoretical ratio, 2.9 \citep{OF}, to $\sim$ 3.1, for broad-line decrements \citep{XD2008}. Since our sample consists of QSO2s, we are only detecting Balmer lines from the NLR, therefore, to calculate the reddening, we assume an intrinsic ratio of 3.0. Taking into account this intrinsic ratio, and the Galactic extinction curve \citep{SM1979}\footnote[3]{Using the extinction curve of \citet{C1989}, results in a less than 4\% change in the computed reddening, compared to the curve of \citet{SM1979}.}, we calculate the reddening, \textit{E(B-V)}, for all the targets in our sample. We assume that the same reddening applies to the entire [O~III] emitting regions for each target in our sample. Note that the extinction occurs in dust along our line of sight, both within the Galaxy and the individual QSOs. The results are listed in Table \ref{tab:Table_2}. \par 
To calculate the values for the bolometric luminosity, $L_{bol}$, for our targets, we correct the values of $L_{[O~III]}$ as described in \citet{S1979}, using the reddening listed in Table \ref{tab:Table_2}. Then, we calculate $L_{bol}$ using the corrected $L_{[O~III]}$ and the bolometric correction factor from \citet{L2009}, which is 454 for the range in luminosities in our sample\footnote[4]{There are other ways to calculate the bolometric luminosity using the combined [O~III]+[O~I] or H$\beta$ and [O~III], as described in \citet{Ne2009}. Based on the SSD spectra, using these methods, we obtain $L_{bol}=1.4\times10^{45}$ erg~${\rm s^{-1}}$ for Mrk 34, which is in reasonable agreement with what we obtain using the corrected [O~III] and the Lamastra correction for this target.}. Our results are listed in Table \ref{tab:Table_2}. \par 

  For the entire Spectral Energy Distribution (SED), we assume that it can be fitted using a number of broken power-laws of the form:
 
\begin{equation}
    L_{\nu} \propto \nu^{-\alpha}
\end{equation}

\noindent where $ \alpha$, the spectral or energy index, is a positive number \citep[e.g.,][]{L1997, M2011}. We assume that the UV to lower energy, “soft” X-ray, is characterised by one value of $ \alpha$, while the higher energy, “hard” X-ray, has a lower value of $ \alpha$. For our study we adopt a cutoff at 100 keV and we set the breakpoint at 500 eV, using the following values \citep{R2018}: 
\medskip

 $ \alpha$ = 1.0 for $h \nu$ $<$ 13.6 eV;  \par
 $ \alpha$ = 1.4 for 13.6 eV $\leq$ ${ h \nu}$ $\leq$ 500 eV;  \par
 $ \alpha$ = 1.0 for 500 eV $\leq$ ${ h \nu}$ $\leq$ 10 keV; \par 
 $ \alpha$ = 0.5 for 10 keV $\leq$ ${ h \nu}$ $\leq$ 100 keV; \par
\noindent It is likely that the "soft" X-ray continuum is more complex than what we assume \citep[e.g.,][]{Ne2002, Kr2002}. However, we opt to use the same SED as in \citet{R2018} to allow for a direct comparison of the results for Mrk 34.\par 

The number of ionising photons per second emitted by the AGN, based on this SED, is:

\small 
\begin{equation}
 Q =  C_{2} \int_{13.6 eV}^{500 eV}(\frac{\nu^{-1.4}}{h\nu}) d\nu + C_{1} \int_{500 eV}^{10 keV} (\frac{\nu^{-1.0}}{h\nu}) d\nu + C \int_{10 keV}^{100 keV} (\frac{\nu^{-0.5}}{h\nu}) d\nu 
\end{equation}
\normalsize

\noindent where $h$ is the Planck constant.\par 
Specifically for Mrk 34, \citet{Ga2014} were able to determine the X-ray luminosity, $L_{(2-10keV)} = 9(\pm 3) \times 10^{43}$ ${\rm erg. s^{-1}}$, which makes it possible to calculate Q for this target. They did not detect any significant variability between their \textit{NuSTAR} and \textit{XMM-Newton} observations. Therefore, we do not consider the possible variability of the X-ray source. The constants $C$, $C_{1}$ and $C_{2}$ were determined by normalizing to the $L_{(2-10keV)}$. \par 
We then use the corrected $L_{[O~III]}$ for Mrk 34 to get the ratio $Q/L_{[O~III]}$, which we apply to all the other QSO2s. Our results are presented in Table \ref{tab:Table_2}.\par
In order to estimate the uncertainties introduced by our SED, we recalculate Q for $1.0 \leq \alpha \leq 1.8$, over the range from 13.6 eV to 500 eV. This results in a factor of 3 change compared to the value of Q computed for $\alpha$ = 1.4. Other sources of uncertainty include those of: the X-ray luminosity, which \citeauthor{Ga2014} assumed to be $\sim$ 30\%; the uncertainty in the ratio $H\alpha$/$H\beta$, $\sim$ 15\%; a factor of 2 in our assumed ionization parameter; the deprojected positions, $\sim$ 10\%; and a factor of 2 uncertainty in the correction for the bolometric luminosity \citep{L2009}. Adding all these in quadrature, results in an uncertainty of a factor of $\sim 4$, which applies to the hydrogen density determination (see section \ref{sec:sec_2.5}).\par 
We also can compare the value for $L_{bol}$ for Mrk 34 derived from the corrected [O~III] with the value computed from our model SED. The latter method consists of extending the SED down to 1 eV and, using $\alpha=1.0$ from 1 eV to 13.6 eV, calculating the bolometric luminosity by integrating over the continuum. The value we get for $L_{bol}$ from our SED is $\sim$ 58\% of the [O~III] derived value. The difference could be due to our assumed SED or the reddening correction applied to the [O~III] emission. However, the difference is less than the uncertainties in Q discussed in the previous paragraph.\par

\subsection{Photoionisation Models and Constraints on Gas Density}
\label{sec:sec_2.5} 

To convert observed [O~III] fluxes in a slit element into masses of gas at that position we need to know the volume of emitting gas and its density. This is conventionally done by using
density-sensitive line ratios such as the [S~II] and [O~II] doublet ratios. Since we only have observations of [O~III] (and $H\beta$ for three of the targets in our sample, as discussed in section \ref{sec:sec_2.4}), we estimate the density from the dimensionless ionization parameter, U, where:
  
  \begin{equation}
      U=\frac{Q}{4\pi r^{2}n_{H}c}
  \end{equation}
 \noindent where $r$ is the radial distance from the AGN, $ n_{H}$ is the hydrogen number density (see section \ref{sec:sec_2.4}) and $c$ is the speed of light. We compute the values of $Q(H)$ for all the QSO2s in the sample, as shown in Table \ref{tab:Table_2}. \par 

\begin{figure}
  \centering
  \includegraphics[width=0.5\textwidth]{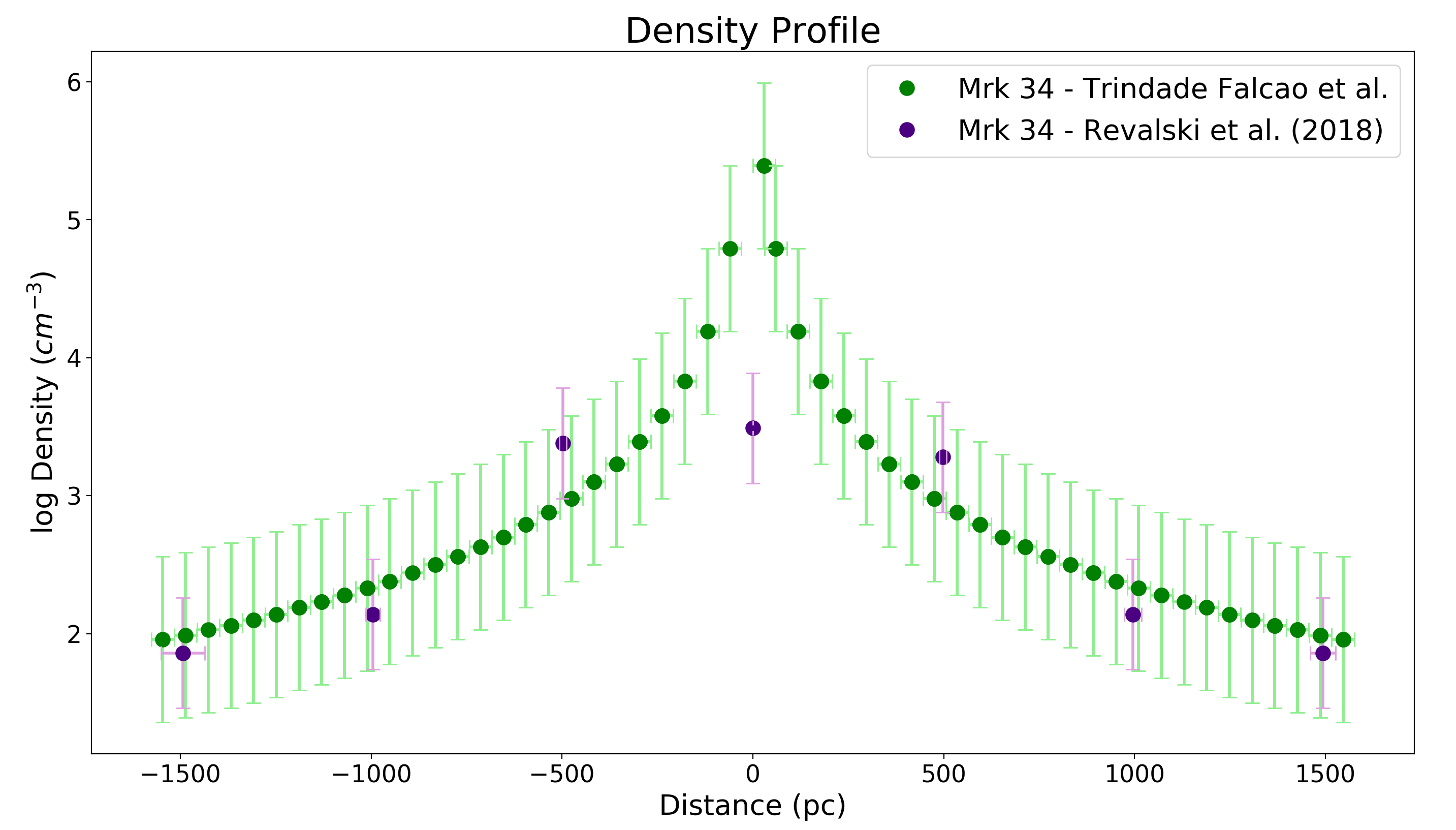}
\caption{\label{fig:Fig_3}A comparison between the gas densities obtained applying our photoionisation models (in green) and the gas densities obtained by Revalski et al. (2018) (in purple) for Mrk 34. The points in purple are the densities for Revalski et al's. medium component. Their point at the centre sums up a region $\sim$ 10 times larger than the region we are sampling. Therefore their density is an average over the sampled region, hence much lower than our computed value.}
\end{figure}

\citet{R2018} used a multi-component photoionisation model for Mrk 34. Based on the model parameters and results given in their Tables 6 and 7, the "medium" component, for which they assumed log($U$) = -2, accounts for most the [O~III] emission. Also, this component contains almost all the mass at each of the points modelled by them, except for 2 positions, where the ionisation is dominated by a higher ionisation component. This results in their determination of higher densities for the [O~III] component, as shown in Figure \ref{fig:Fig_3}. Therefore we model the [O~III] emission-line gas, at each radial position, with a single component of log($U$) = -2. This means that the density drops with $r^{-2}$, for the [O~III] gas along the whole emitting region (as also proposed by \citealt{D2020}).\par 
 We use version 17.00 of Cloudy \citep{Fe2017} to construct photoionisation models with log($U$) = -2. These predict $\frac{[O~III]}{H\beta}$ $\geq$ 10, which we confirmed for the targets with both [O~III] and H$\beta$ based on measurements of the STIS spectra (2MASX J07594101+5050245, 2MASX J11001238+0846157 and 2MASX J16531506+2349431, which have an average $\frac{[O~III]}{H\beta}$ = 10.2).\par
 In dusty gas, emission lines are suppressed by two mechanisms. One is the dust absorption of multiply scattered resonance lines, such as Ly-$\alpha$ \citep[e.g.,][]{K1986}. The other is due to the depletion of elements, such as Si, Mg, and Fe onto dust grains. STIS long-slit UV spectra of the NLR typically show fairly strong Mg II $\lambda$2800 \citep{K2000b, K2001, C2005}, which suggests that the refractory elements are not heavily depleted. While these spectra reveal a wide range of Ly-$\alpha$/H$\beta$ ratios, they generally range from 10 - 20, which indicates the presence of some dust within the emission-line gas. Based on these studies, we assume a dust-to-gas ratio of 50\% that of the Galactic interstellar medium with proportional depletion of elements from gas phase, consistent with \citet{R2018}. This dust is not the source of the line of sight reddening discussed in section \ref{sec:sec_2.4}.\par 
While super-solar abundances appear to be common in broad line region gas, at least in QSOs \citep[e.g.,][]{HF1998, D2003}, NLR studies indicate abundances closer to solar \citep{N2001, N2002}. Based on their photoionisation analysis of the NLR, \citet{Gr2004} argued for a N/H ratio approximately twice solar, consistent with other NLR studies \citep[e.g.,][]{ST1996,ST1998,K1998}. Since N/H increases as $(Z/Z_{solar})^2$ \citep{TA1973}, where $Z$ is the fractional abundance of heavy elements, this ratio corresponds to elemental abundances of approximately 1.4x solar, which we adopt for this study. \par 

 The exact logarithmic values relative to hydrogen by number are:  ${\rm C=-3.54}$, ${\rm N=-3.88}$, ${\rm O=-3.205}$, ${\rm Ne=-3.92}$, ${\rm Na=-5.61}$, ${\rm Mg=-4.47}$, ${\rm Al=-5.70}$, ${\rm Si =-4.64}$, ${\rm P=-6.44}$, ${\rm S=-4.73}$, ${\rm Ar=-5.45}$, ${\rm Ca=-5.81}$, ${\rm Fe=-4.65}$, ${\rm Ni=-5.93}$. In order to account for the grain composition, we include depletions from gas phase for C, O and the refractory elements \citep[e.g.,][]{SW1996}. Note that dust can play an important role in the dynamics of outflows \citep[e.g.,][]{BN2019}. We will be examining the outflow dynamics in our subsequent paper (Trindade Falcão et al. in preparation). \par

For reference, we calculate the maximum possible column density of the gas emitting [O~III]. This is when the gas is radiation bounded. Specifically, the modelling integration stopped when the electron temperature drops below 4000K, at which point the ionising radiation has essentially been exhausted. It is possible that the total column densities are greater than this, but we have no means of detecting that gas in these data. On the other hand, it is quite likely that the gas is not radiation bounded in places. What matters is the volume of gas emitting [O~III]. We calculate this in the next section.\par 
As shown in Figure \ref{fig:Fig_3}, this method does a reasonable job in matching densities from detailed photoionisation models. We then use the computed column densities, $N_H$, to calculate the masses and, subsequently, the other physical properties of the [O~III] gas, as described in section \ref{sec:sec_4}.\par

\section{Calculations}
\label{sec:sec_3}

\begin{figure*}
  \centering
 \begin{minipage}[b]{0.45\textwidth}
  \includegraphics[width=8.65cm]{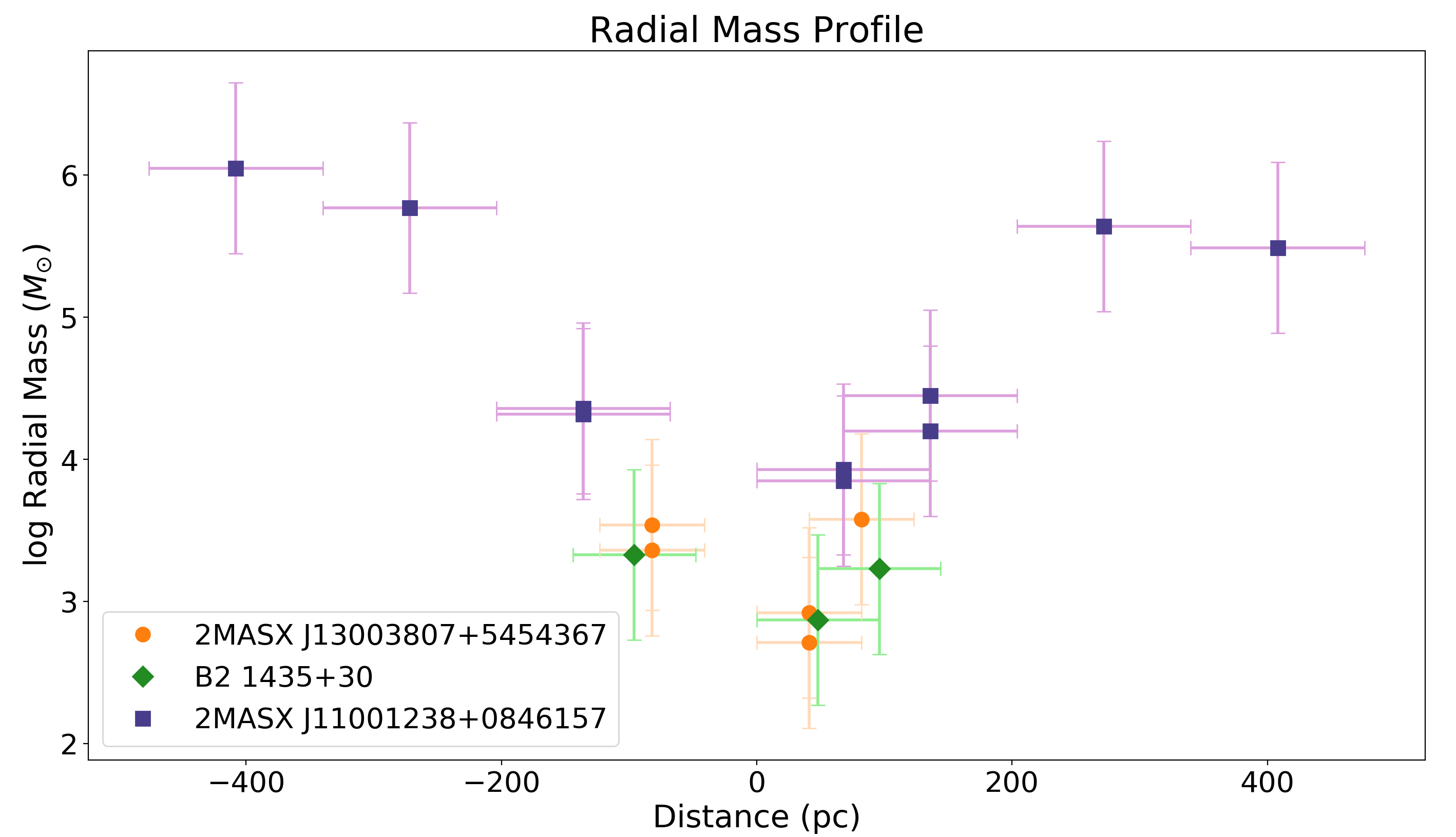}
 \end{minipage}\qquad 
 \begin{minipage}[b]{0.45\textwidth}
  \includegraphics[width=8.65cm]{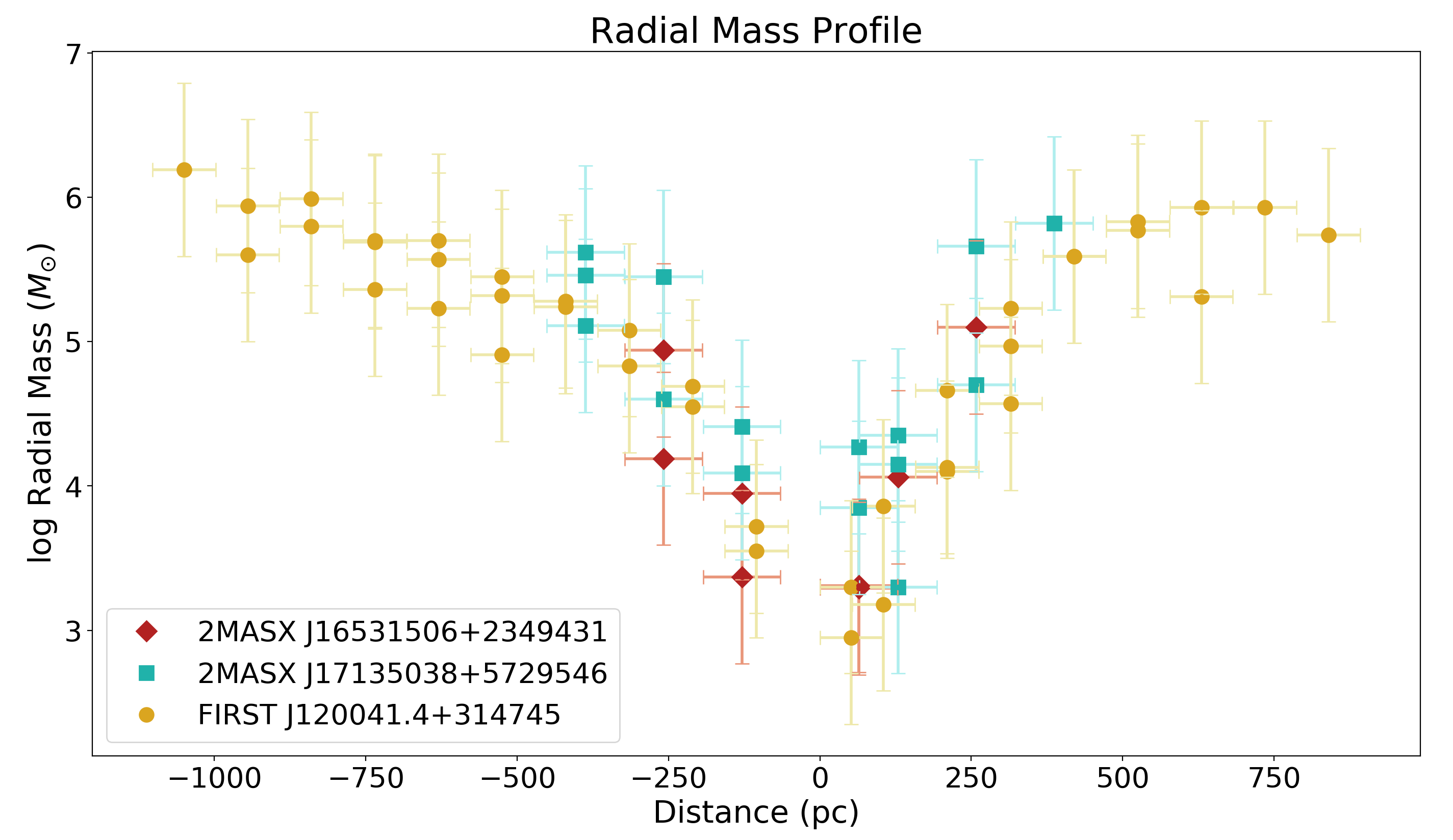}
 \end{minipage}\qquad 
 \begin{minipage}[b]{0.45\textwidth}
  \includegraphics[width=8.65cm]{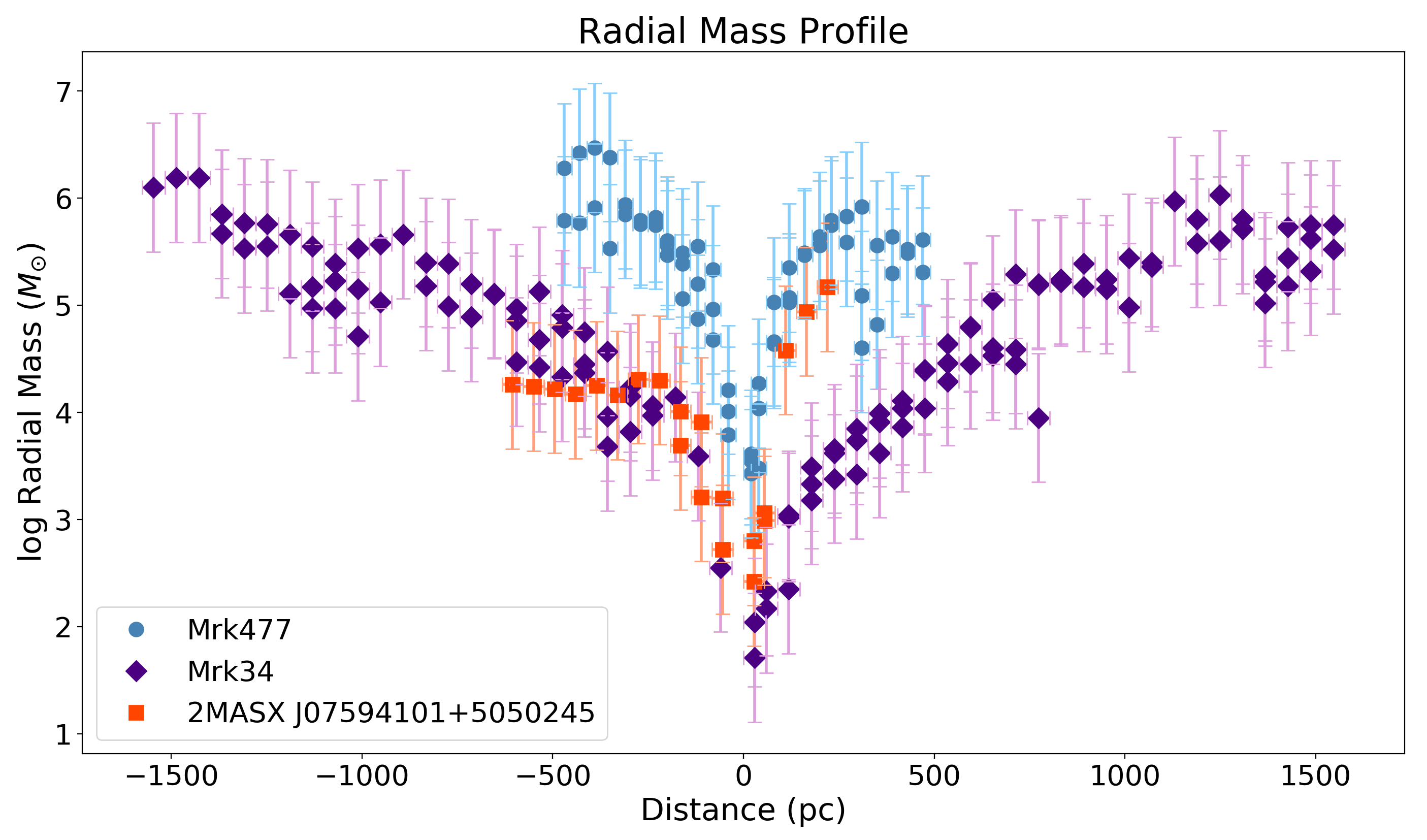}
\end{minipage}\qquad 
 \begin{minipage}[b]{0.45\textwidth}
  \includegraphics[width=8.65cm]{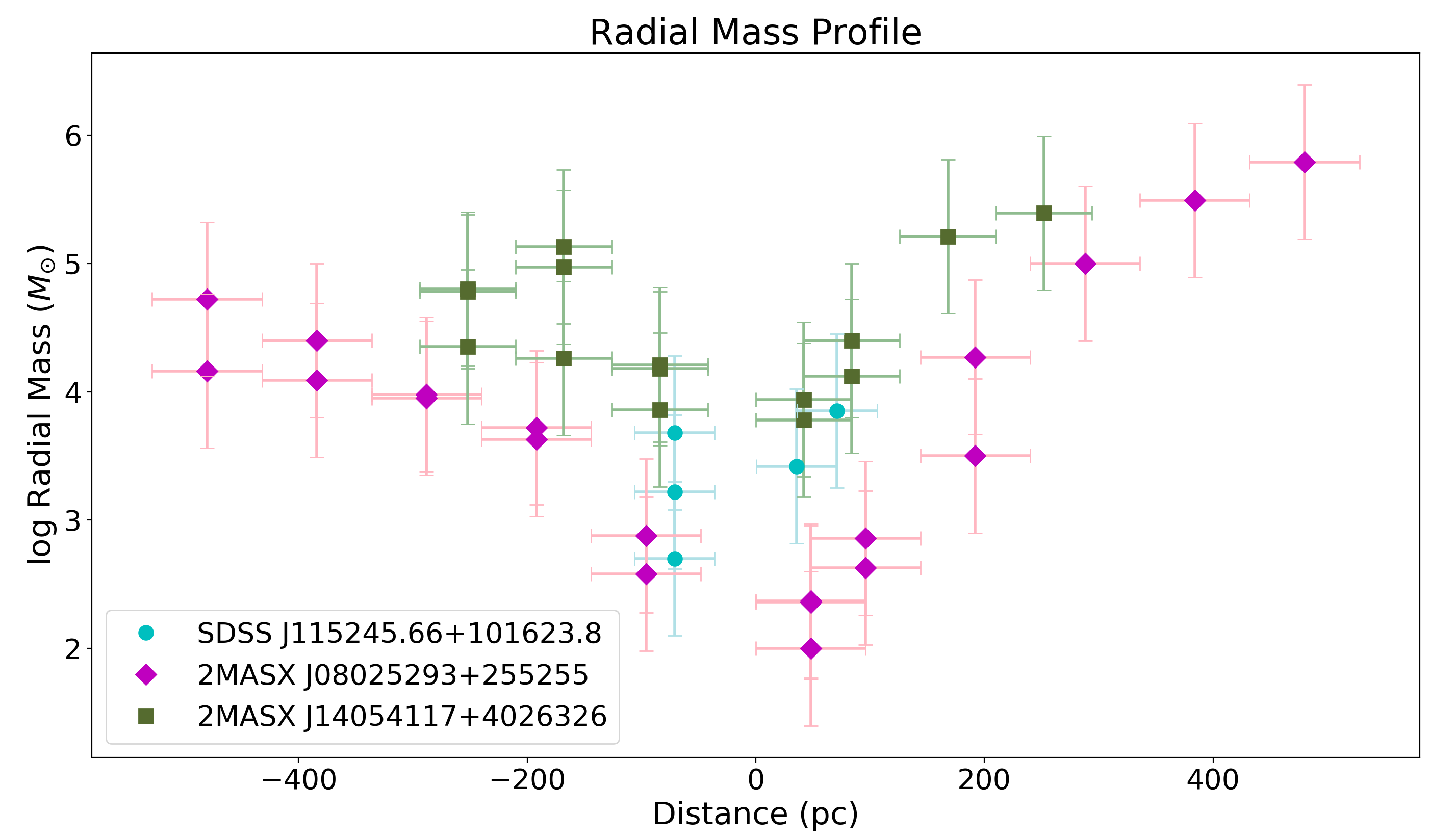}
 \end{minipage}\qquad 
\caption{ Ionised mass profile in units of $M\textsubscript{\(\odot\)}$ calculated from the total flux in each semi-elliptical annulus for all the targets in our sample. QSO2s with similar redshifts were plotted together. All points considered in the plots are inside the range of outflow defined in Fischer et al. 2018. With a few exceptions, most mass profiles look similar. So the total outflow mass is primarily determined by where the outflow stops and the rotation kinematics begin. Each annulus has a $\delta$r = 0.05 arcsec, and, due to the range in distances of our targets, $\delta$r corresponds to a range of physical lengths in pc.}
\label{fig:Fig_4}
\end{figure*}

To determine the various quantities associated with the outflow as a function of distance from the SMBH we first need to determine the mass of gas in each of our semi-annuli. We do this by first estimating $M_{slit}$, the mass of gas inside the \textit{HST} STIS slit at a given radius, and then scaling this to the mass in the whole semi-annulus by multiplying the ratio of the total [O~III] flux in the semi-annulus, ${{F_{[O III]}}_{ann}}$, to the flux in the slit, ${{F_{[O III]}}_{m}}$ . For our calculations we only consider points lying inside the range of outflow defined in \citetalias{F2018}.\par 
The mass of gas emitting [O~III], in grams, in each position along the slit is giving by \citep{C2015}:

\begin{equation}
    M_{slit}(r) = N_H\mu m_p \left(\frac{L_{[O~III]}}{{F_{[O~III]}}_{c}}\right)
\end{equation}
\noindent where $N_H$ is the hydrogen column density, assumed to be the same as the column density modelled by Cloudy,  $\mu$ is the mean mass per proton\footnote[5]{We use $\mu$ = 1.4, which is consistent with roughly solar abundances.}, $m_p$ is the mass of a proton. To get the mass of gas the column density needs to be multiplied by an
effective area. The term in parentheses gives the effective surface area of the emitting gas as seen by the observer. ${{F_{[O~III]}}_{c}}$ is the [O~III] luminosity per ${\rm cm^{2}}$ calculated by Cloudy and
$L_{[O~III]}$ is the observed luminosity calculated from the reddening corrected flux in the slit. That is, 

\begin{equation}
    L_{[O~III]} = 4\pi D^2 {{F_{[O~III]}}_{m}}
\end{equation}

\noindent where $D$ is the distance to the QSO2s (see Table \ref{tab:Table_1}) and ${{F_{[O~III]}}_{m}}$ is the intrinsic flux measure at each point in the STIS spectra.\par
Physically, the equation for $M_{slit}$ (Equation 4) determines the area of the emitting clouds through the ratio of the luminosity and flux, and then multiplies this by the column density to yield the total number of particles, which, when multiplied by the mean mass per particle, gives the total ionised mass.\par
We estimate the mass of gas in the half annulus at a given radial distance from the centre by scaling the mass in the slit (Equation 4) by the ratio of the flux
in the entire semi-annulus to the flux in the slit.

Specifically, the total ionised mass in outflow at each radial distance is: 

 \begin{equation}
     M_{out}(r)=M_{slit}(r) \left(\frac{{F_{[O III]}}_{ann}}{{F_{[O~III]}}_{m}}\right)
 \end{equation}

 \noindent where ${{F_{[O III]}}_{ann}}$ is the flux in each half image annuli of width $\delta r$, as shown in Figure \ref{fig:Fig_2}. Thus, our method
assumes that throughout the semi-annulus the gas has the same density, $N_{H}$ , and outflow velocity, $v_{out}$, as the gas at the slit location.\par
 After calculating $M_{out}(r)$, we are able to estimate the mass outflow rates ($ \dot M_{out}(r)$), kinetic energies ($E$), kinetic luminosities ($\dot E$), momenta ($p$) and momentum flow rates ($\dot p$). All these quantities are related to the power and impact of the NLR outflows \citep{KP2015}. \par 
 The mass outflow rates ($\dot M_{out}(r)$) are calculated, at each point along the NLR, using:

\begin{equation}
    \dot M_{out}(r) = \frac{M_{out}(r) v_{out}}{\delta r} 
\end{equation}

\noindent where $v_{out}$ is the deprojected outflow velocity at the distance of the semi-annulus; the deprojection factors are the same as those used in \citetalias{F2018} (see section \ref{sec:sec_2.3}).\par 
The kinetic energies (in ergs), kinetic luminosities (in erg/s), momenta (in dyne-s) and momentum flow rates (in dyne\footnote[6]{1 dyne = 1 g~cm~${\rm s^{-2}}$ = ${\rm 10^{-5}}$ N}) are given by:

\begin{equation}
    E(r) = \frac{1}{2} M_{out} v_{out}^2 
\end{equation}

\begin{equation}
    \dot E(r) = \frac{1}{2} \dot M_{out} v_{out}^2      
\end{equation}

\begin{equation}
    p(r) = M_{out}v_{out}
\end{equation}

\begin{equation}
    \dot p(r) = \dot M_{out}v_{out}
\end{equation}

\begin{figure*}
  \centering
 \begin{minipage}[b]{0.45\textwidth}
  \includegraphics[width=8.65cm]{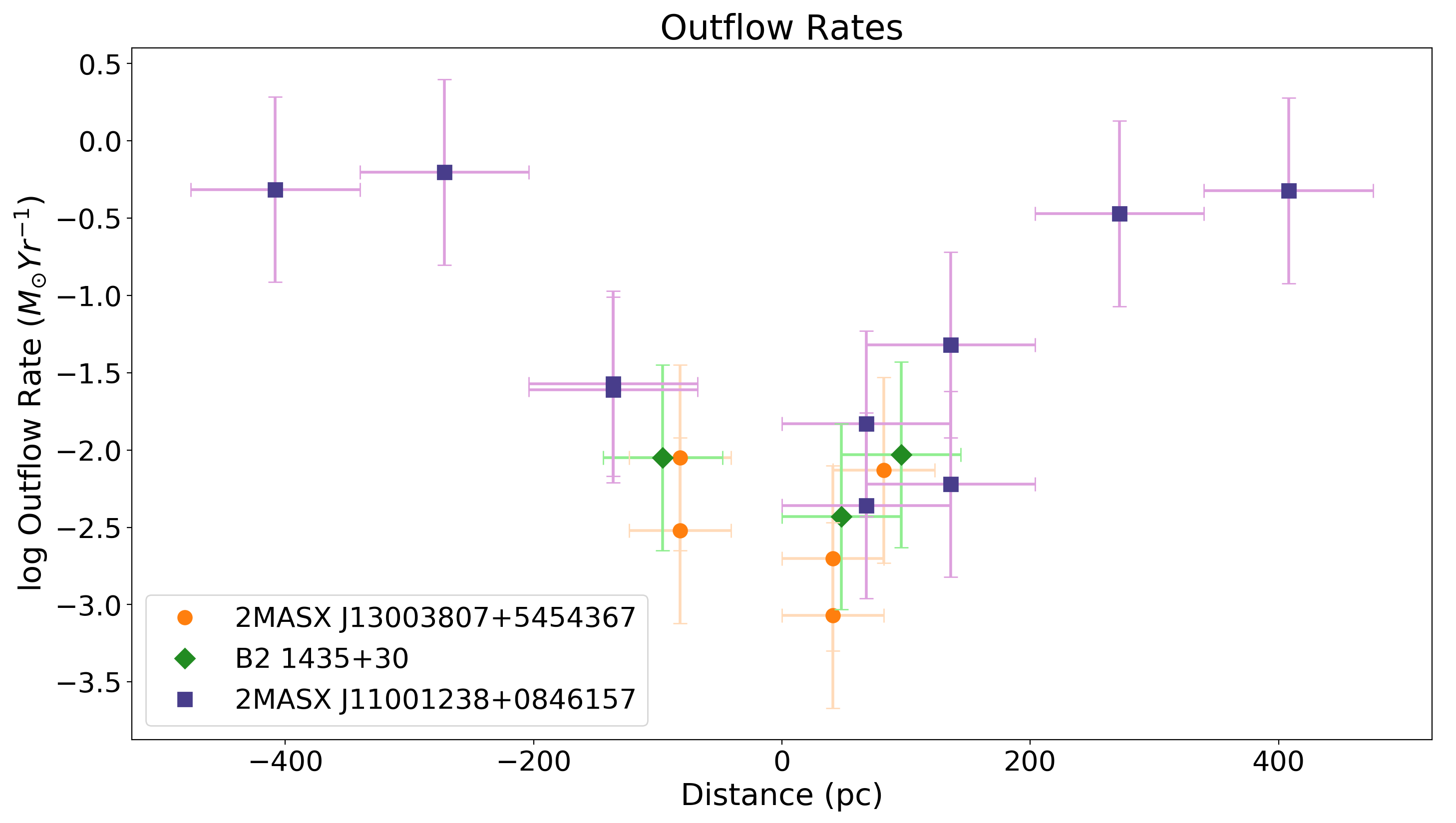}
 \end{minipage}\qquad
 \begin{minipage}[b]{0.45\textwidth}
  \includegraphics[width=8.65cm]{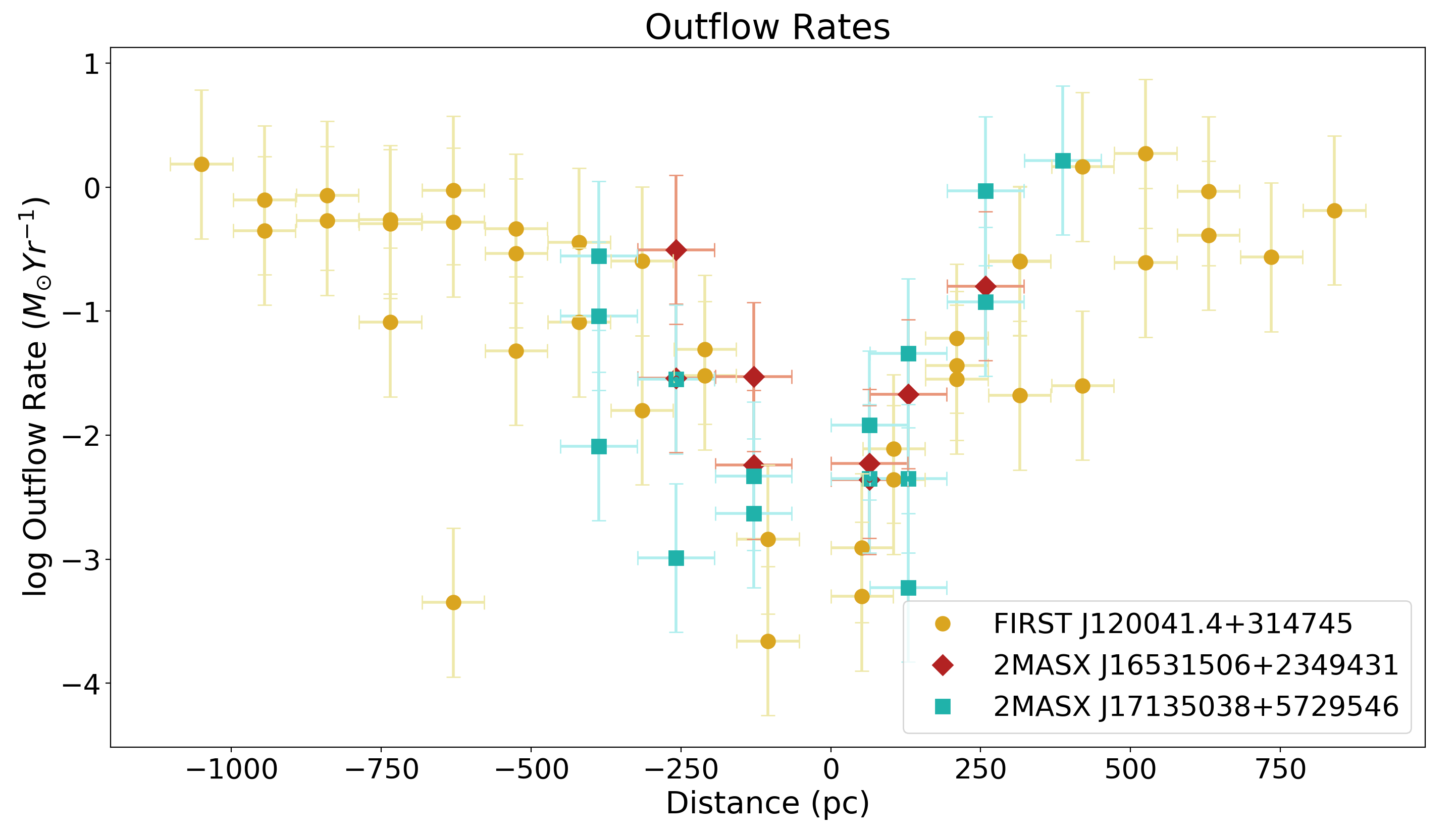}
 \end{minipage}\qquad
 \begin{minipage}[b]{0.45\textwidth}
  \includegraphics[width=8.65cm]{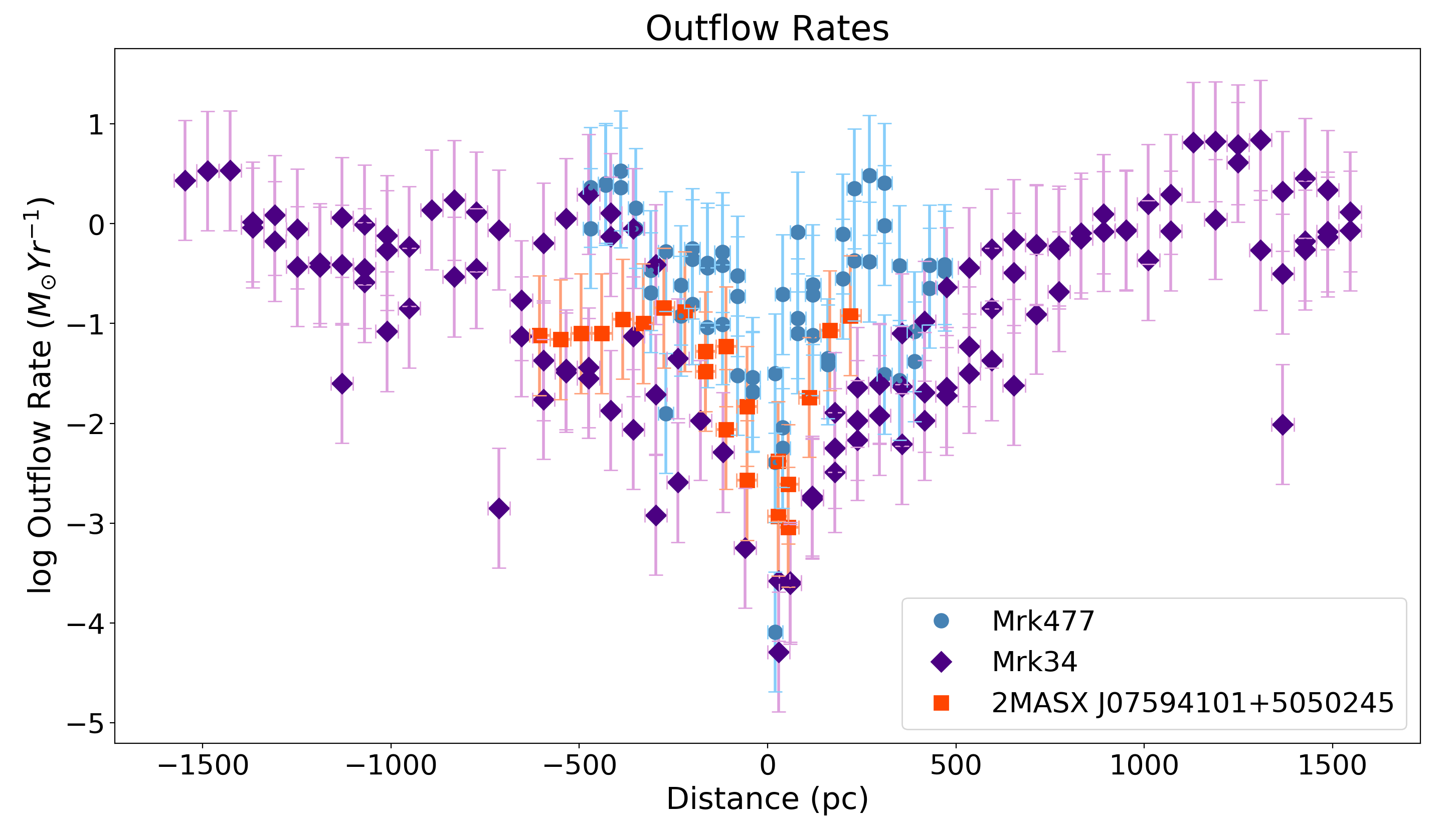}
 \end{minipage}\qquad
 \begin{minipage}[b]{0.45\textwidth}
  \includegraphics[width=8.65cm]{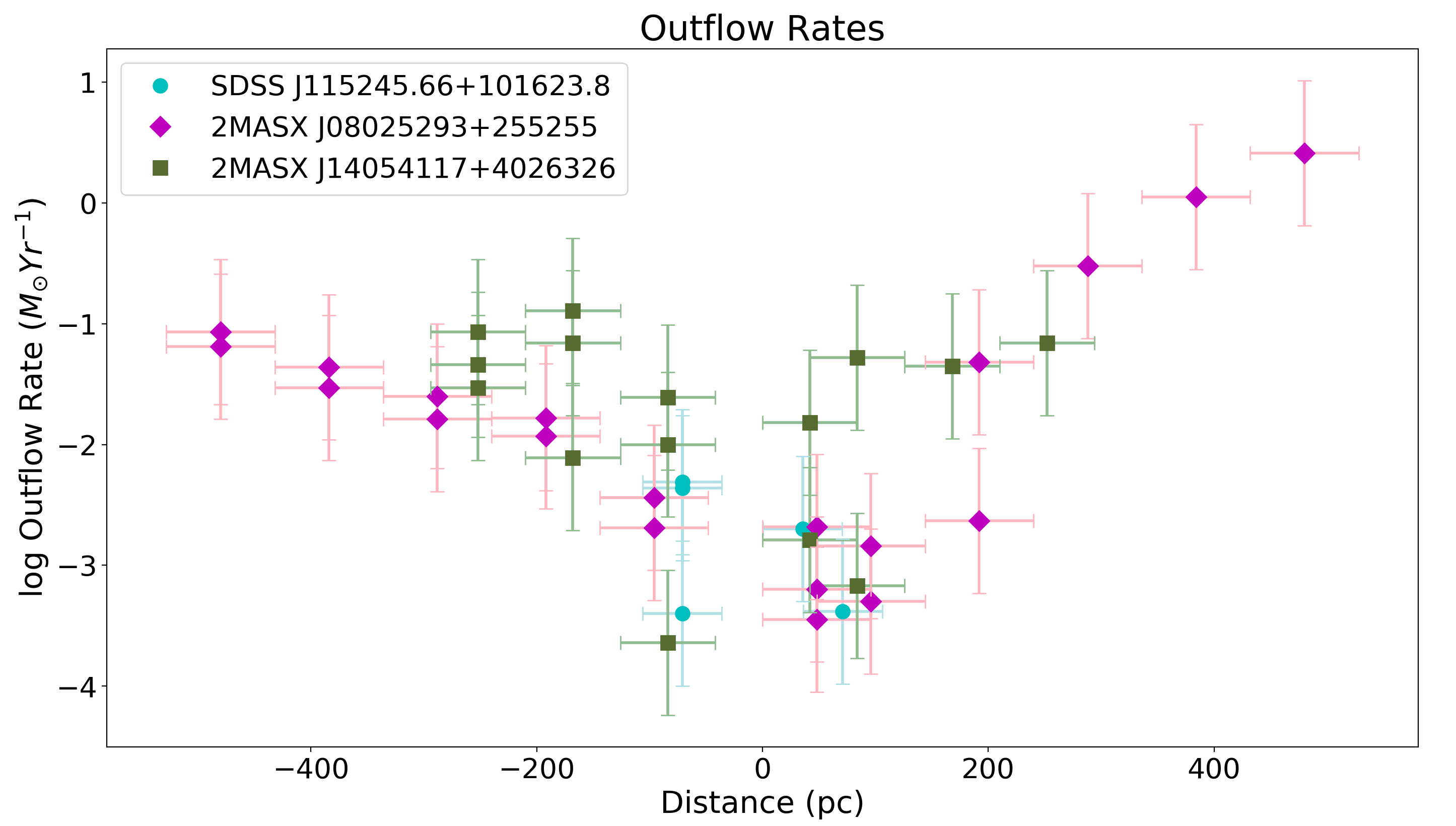}
 \end{minipage}\qquad
\caption{Spatially-resolved outflow rates for all targets, assuming that all of the material is in outflow. Same as Figure \ref{fig:Fig_4}}
\label{fig:Fig_5}
\end{figure*}

The momenta and momentum flow rates can be compared to the AGN bolometric luminosity, and the radiation pressure force, $\frac{L}{c}$, to quantify the efficiency of the NLR in converting radiation from the AGN into the radial motion of the outflows \citep{ZK2012, Co2014}. \par
 
 All the uncertainties discussed in section \ref{sec:sec_2.4}, which affect our determination of the hydrogen density, result in corresponding uncertainty in our mass calculations. Hence the quantities computed in Equations 7 - 11 have the same factors of uncertainty. \par

 \begin{figure*}
  \centering
 \begin{minipage}[b]{0.45\textwidth}
  \includegraphics[width=8.65cm]{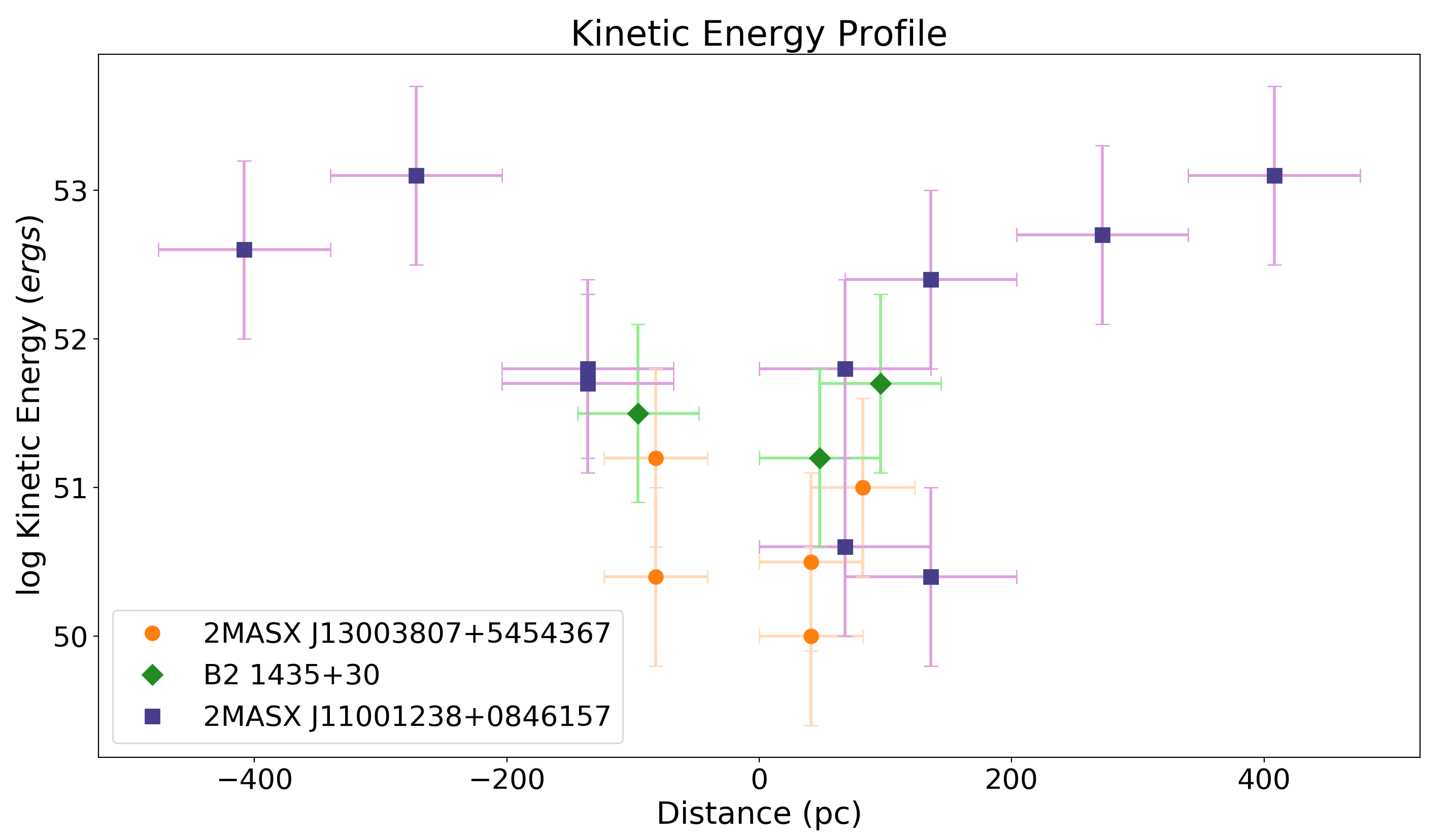}
 \end{minipage}\qquad 
 \begin{minipage}[b]{0.45\textwidth}
  \includegraphics[width=8.65cm]{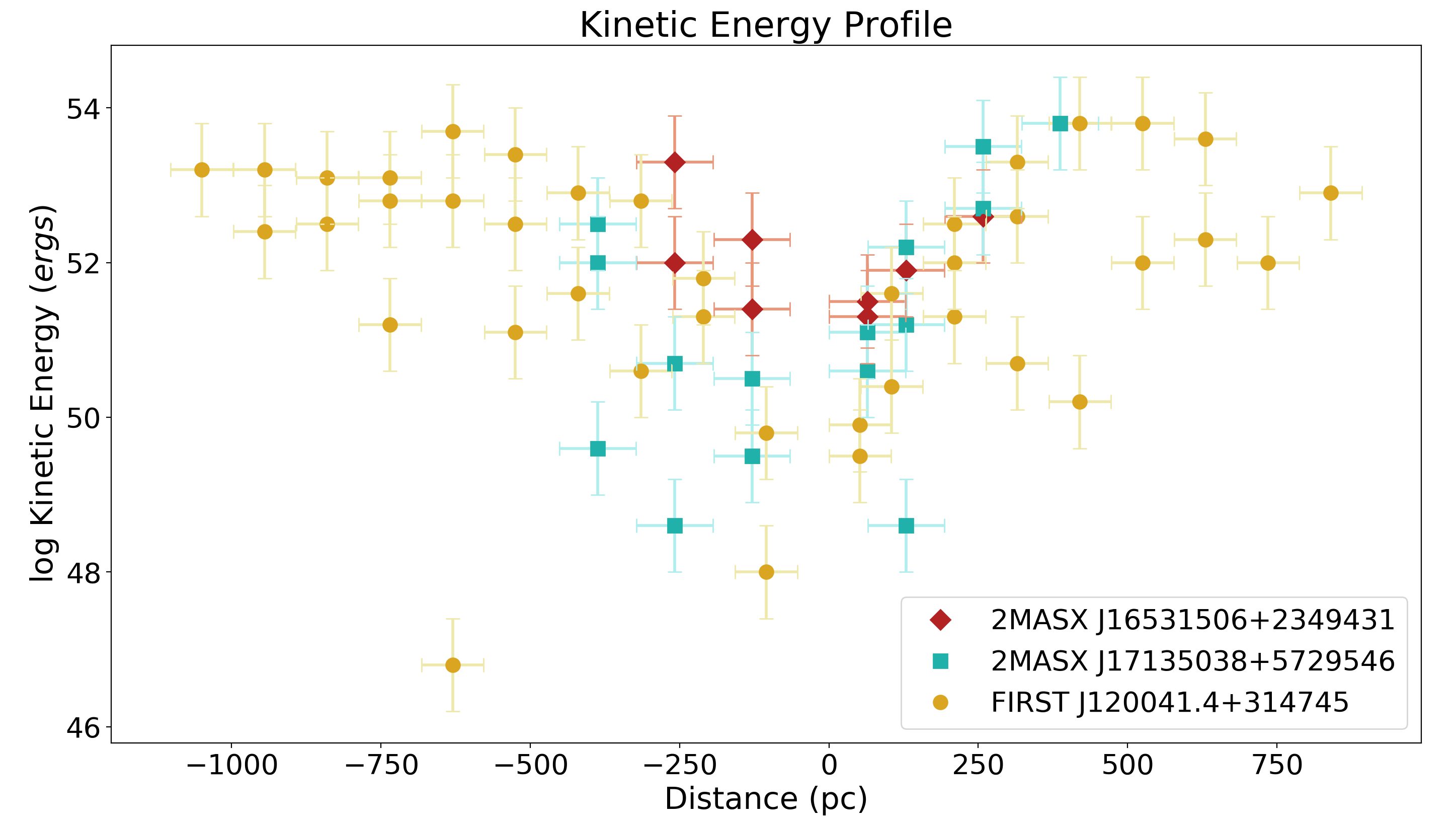}
 \end{minipage}\qquad
 \begin{minipage}[b]{0.45\textwidth}
  \includegraphics[width=8.65cm]{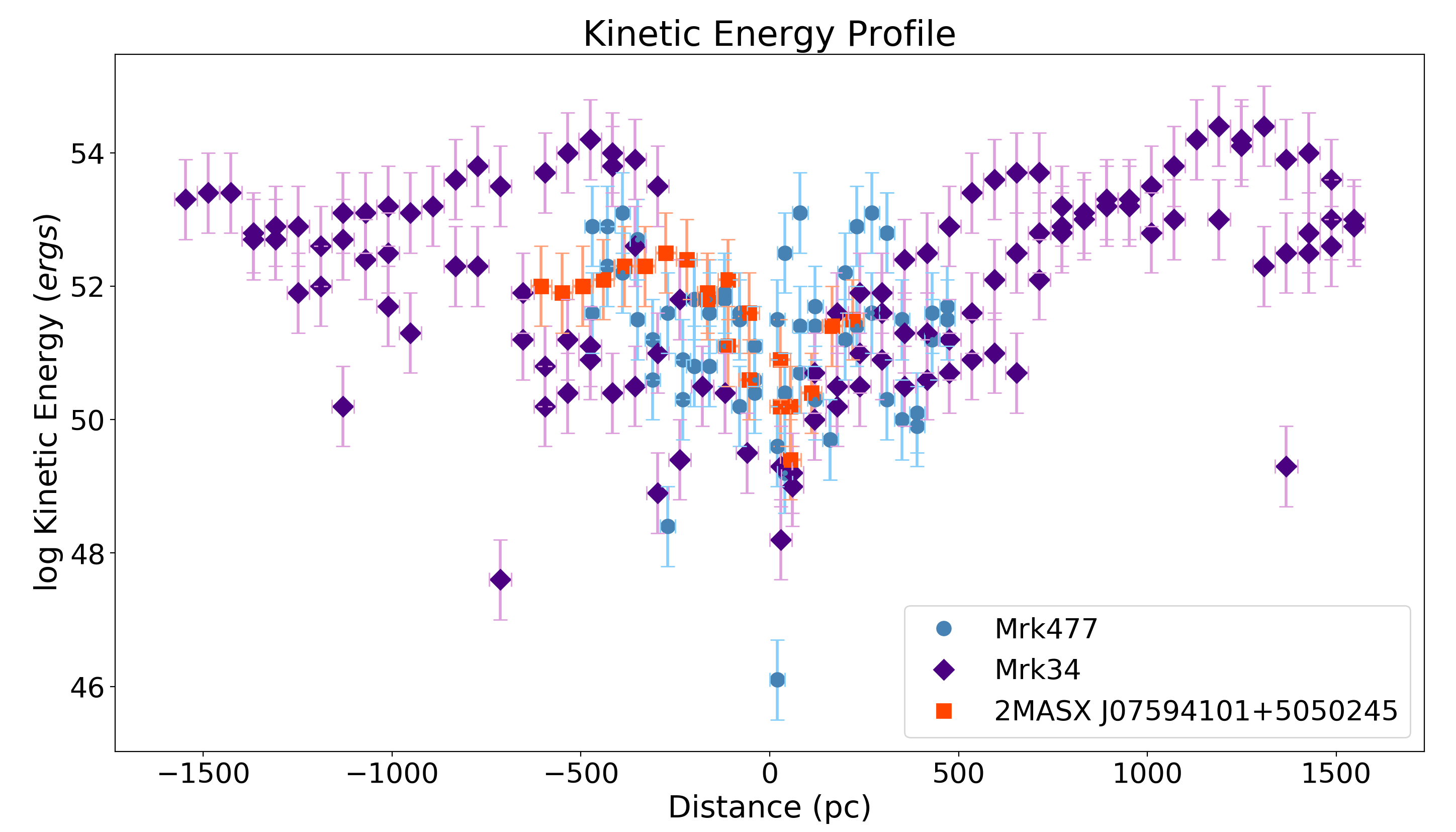}
 \end{minipage}\qquad
 \begin{minipage}[b]{0.45\textwidth}
  \includegraphics[width=8.65cm]{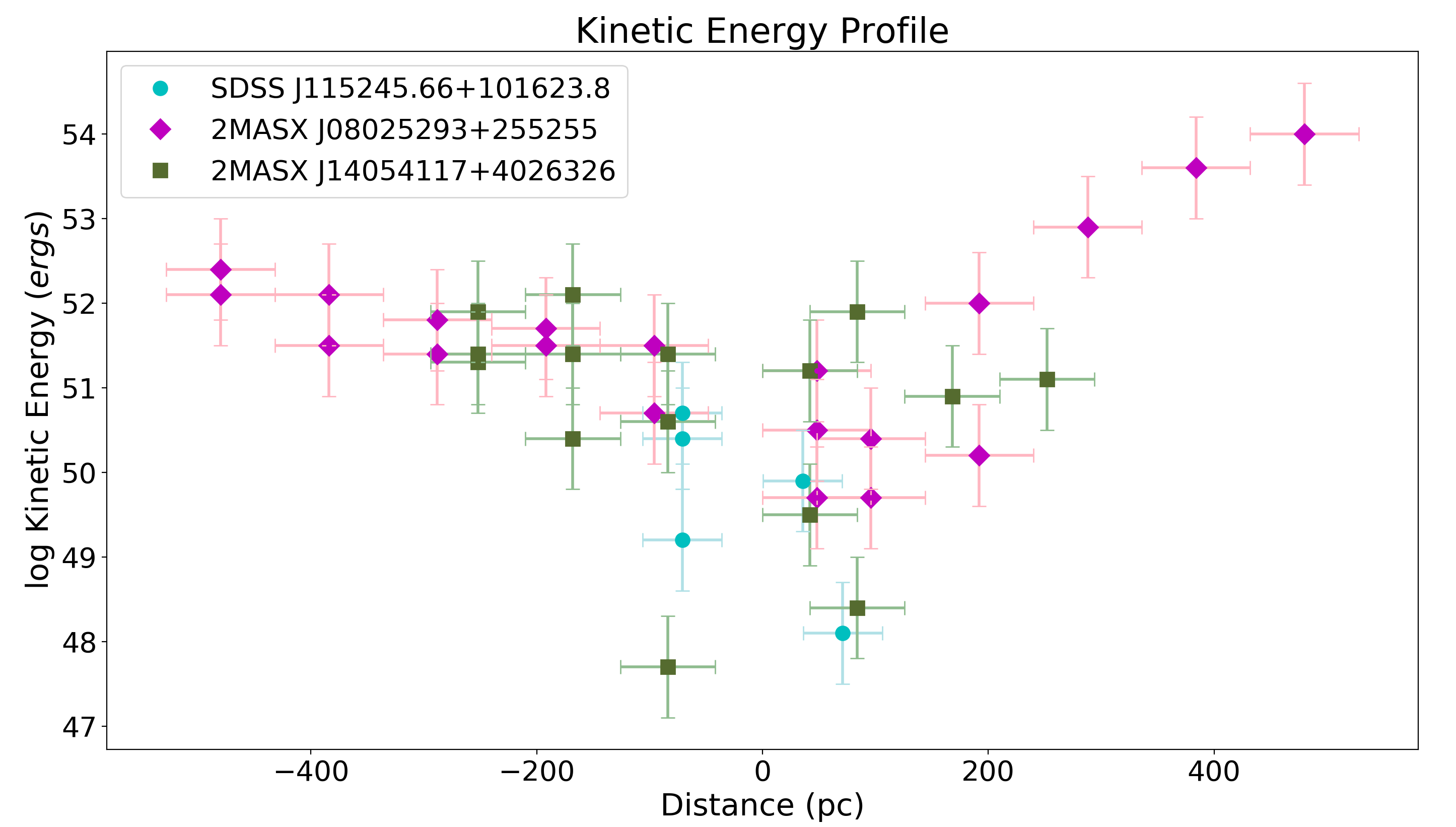}
 \end{minipage}\qquad

\caption{ Kinetic Energy profiles for all targets. See section \ref{sec:sec_4}.}
\label{ref:Fig_6}
\end{figure*}

 \begin{table*}

\caption{Column 4 lists the distance to the QSO2, considering a Hubble constant = 71 ${\rm km}$ ${\rm s^{-1}}$ ${\rm Mpc^{-1}}$. Column 5 lists the maximum radial extent of the AGN driven outflows, with emission lines, which exhibit high centroid velocities and/or multiple emission lines with multiple components. Column 6 lists the maximum radial extent of the disturbed gas, with low centroid velocities and FWHM $>$ 250 km ${\rm s^{-1}}$ (Fischer et al. 2018). Column 7 lists the $\delta r$ for each target.}
\label{tab:Table_1}

\smallskip
\begin{tabular*}{\textwidth}{@{\extracolsep{\fill}}lcccccc}
 Target & Redshift & Scale  & Distance to QSO2 & Deproj. $R_{out}$ & Deproj. $R_{dis}$ & $\delta r$\\
 & & (kpc/")&  (Mpc) & (kpc) & (kpc) & (pc)\\
 (1) & (2) & (3) & (4) & (5) & (6) & (7)\\
SDSS J115245.66+101623.8  & 0.070 & 1.30 & 296 & 0.15 & 1.23 & 0.71\\
MRK 477 & 0.038 & 0.72 & 161 & 0.54 & 0.90 & 0.03 \\
MRK 34 & 0.051 & 0.95 & 215 & 1.89 & 1.89 & 0.06 \\
2MASX J17135038+5729546 & 0.113 & 1.97 & 477 & 0.65 & $>$0.92 & 0.13 \\
2MASX J16531506+2349431 & 0.103 & 1.83 & 435 & 0.57 & 1.16 & 0.14 \\
2MASX J14054117+4026326 & 0.081 & 1.47 & 342 & 0.33 & $>$0.94 & 0.08 \\
2MASX J13003807+5454367 & 0.088 & 1.59 & 372 & 0.16 & 0.16 & 0.08\\
2MASX J11001238+0846157 & 0.101 & 1.80 & 427 & 0.69 & $>$1.51 & 0.14  \\
2MASX J08025293+2552551 & 0.081 & 1.48  & 342 & 0.57 & 0.89 & 0.09 \\
2MASX J07594101+5050245 & 0.054 & 1.02  & 228 & 0.67 & 0.67 & 0.05 \\
FIRST J120041.4+314745 & 0.116 & 2.04 & 490 & 1.07 & $>$1.59 & 0.10 \\
B2 1435+30 & 0.092 & 1.66 & 389 & 0.20 & $>$1.74 & 0.09 \\
\end{tabular*}
\end{table*}

\begin{figure*}
 \centering
 
 \begin{minipage}[b]{0.45\textwidth}
  \includegraphics[width=8.65cm]{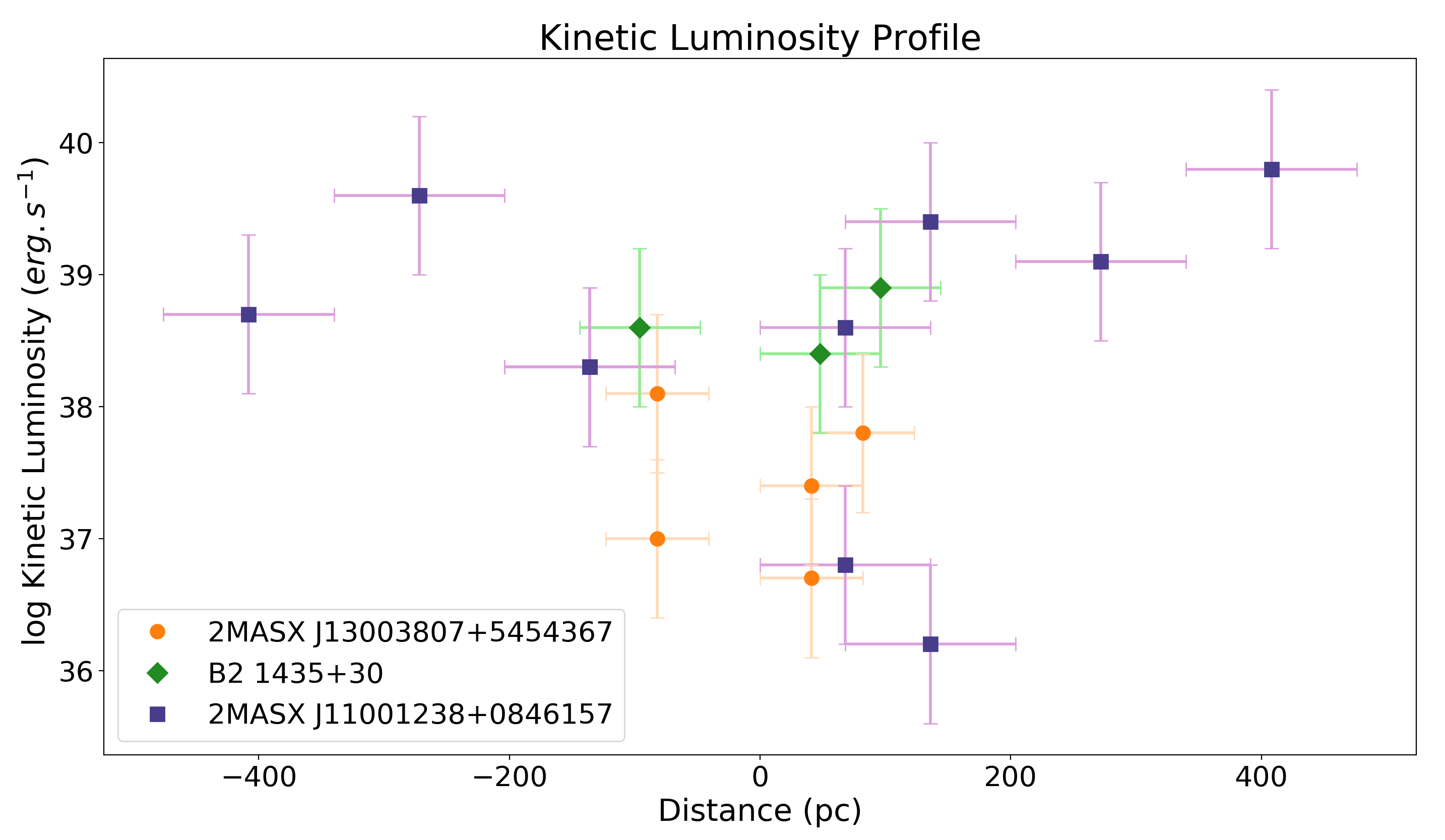}
 \end{minipage}\qquad
 \begin{minipage}[b]{0.45\textwidth}
  \includegraphics[width=8.65cm]{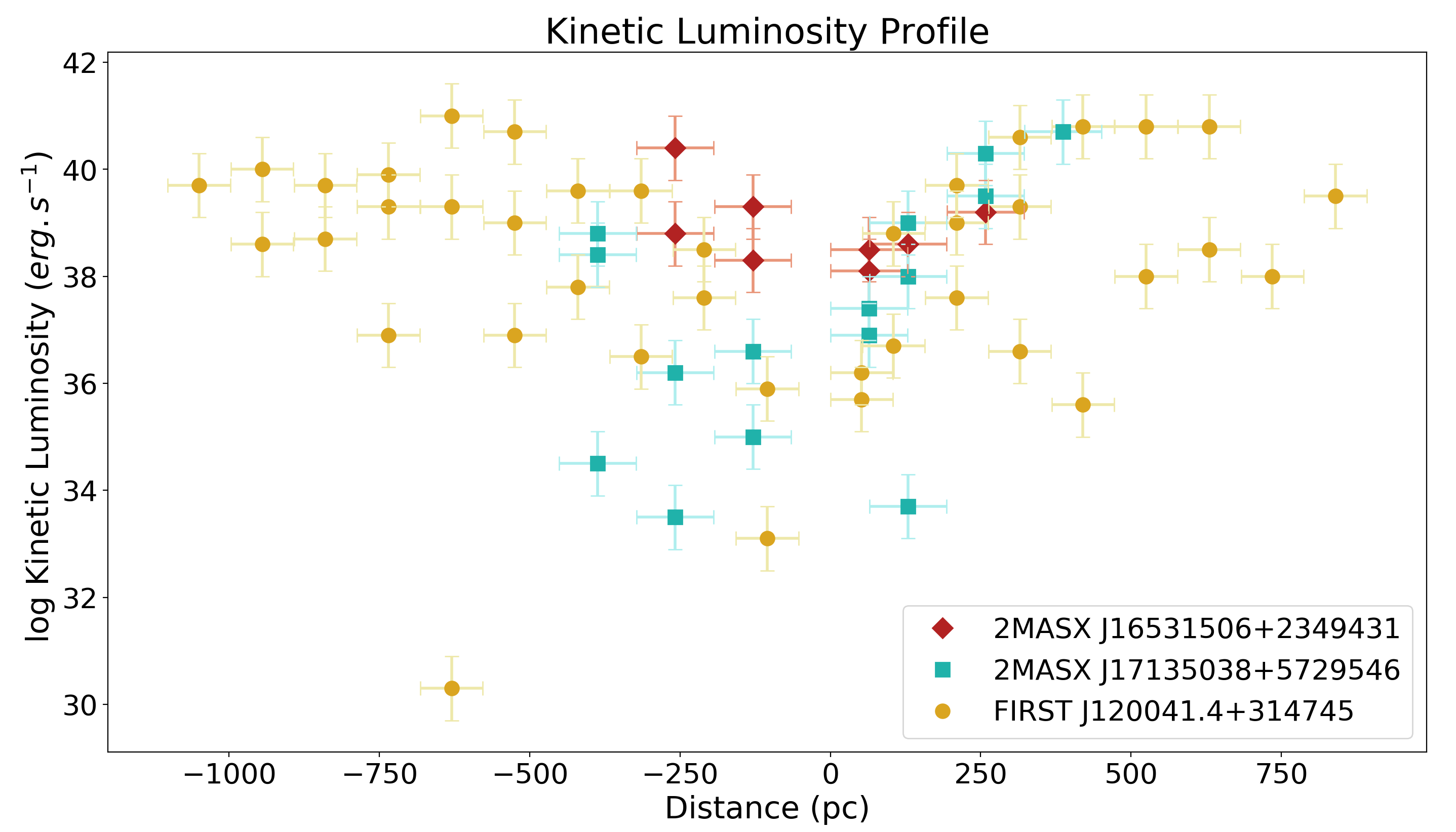}
 \end{minipage}\qquad
 \begin{minipage}[b]{0.45\textwidth}
  \includegraphics[width=8.65cm]{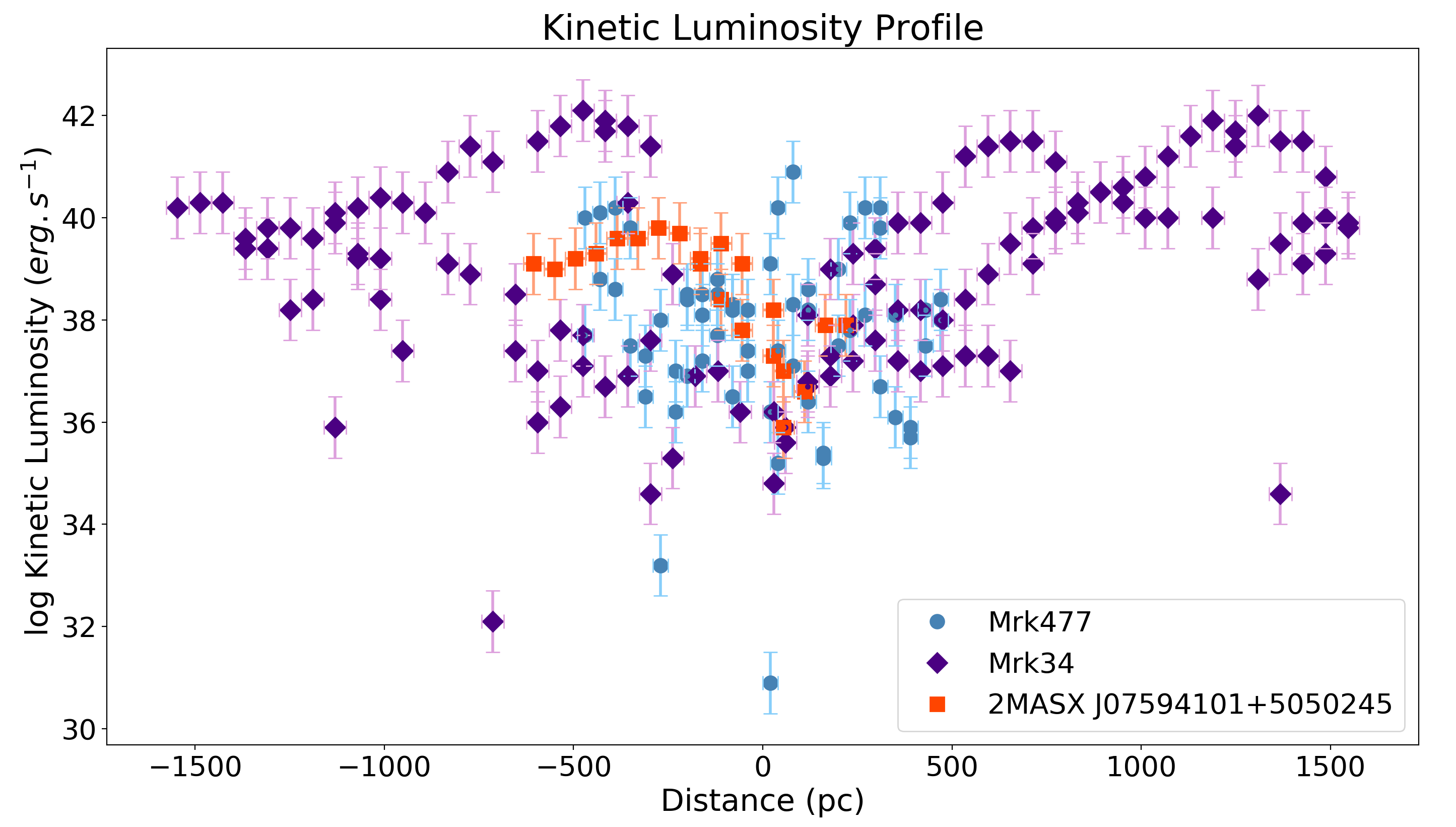}
 \end{minipage}\qquad
 \begin{minipage}[b]{0.45\textwidth}
  \includegraphics[width=8.65cm]{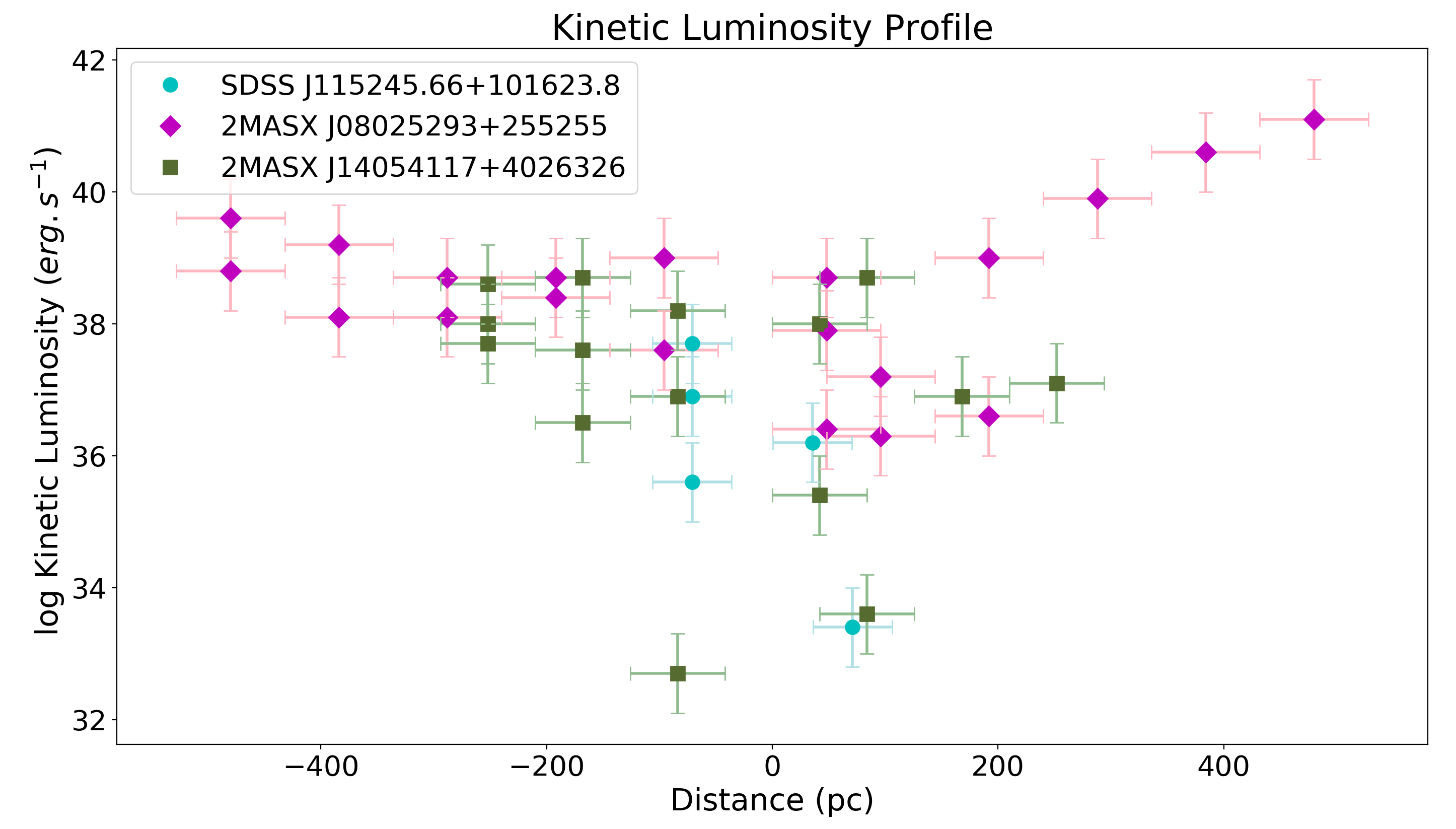}
 \end{minipage}\qquad

\caption{ Kinetic Luminosity Rates calculated for all targets. QSO2s with similar redshifts were plotted together. See section \ref{sec:sec_4}}
\label{fig:Fig_7}

\end{figure*}

\begin{table*}

\caption{Column 2 lists the corrected [O~III] luminosity, as discussed in section \ref{sec:sec_2.4}. All targets have $L_{[O~III]}$ $\geq$ $1.9\times 10^{42}$, satisfying the conventional B-band absolute magnitude criterion of a “quasar”, $B_{mag}$ $<$ -23, where a corresponding $L_{[O~III]}$ is $>$  $3\times 10^8 L\textsubscript{\(\odot\)}$ (Zakamska et al. 2003). Column 3 lists the corrected bolometric luminosity for each QSO2, calculated as described in section \ref{sec:sec_2.4}. Column 4 lists the corrected number of ionising ${\rm photons. s^{-1}}$. Column 5 lists the ratio between the retrieved fluxes of $H\alpha$ and $H\beta$ from SDSS for each target. Column 6 lists the degree of reddening for each target in our sample, calculated using the values for $H\alpha$ and $H\beta$ from their SDSS spectra. Due to the similar redshifts and selection criteria, the range in luminosities is small (by a factor of $\sim$ 4). }
\label{tab:Table_2}

\smallskip
\begin{tabular*}{\textwidth}{@{\extracolsep{\fill}}lcccccr}
  Target & $L_{[OIII]}$ & $L_{bol}$ & Q & $H\alpha$/$H\beta$ &\textit{E(B-V)}\\
   & ${\rm (erg}$ ${\rm s^{-1})}$ &${\rm (erg}$ ${\rm s^{-1})}$ & ${\rm (photons}$ ${\rm s^{-1})}$& & &\\ 
   (1) & (2) & (3) & (4) & (5) & (6) \\
  SDSS J115245.66+101623.8 & $3.5\times10^{42}$ & $1.6\times10^{45}$ & $4.8\times10^{54}$ &3.68 & 0.19 \\
  MRK 477 & $4.0\times10^{42}$ & $1.8\times10^{45}$  & $5.4\times10^{54}$ & 3.76& 0.21  \\
  MRK 34  & $5.7\times10^{42}$ & $2.6\times10^{45}$ & $7.8\times10^{54}$ & 3.98 & 0.26\\
  2MASX J17135038+5729546 & $6.1\times10^{42}$ & $2.8\times10^{45}$  & $8.3\times10^{54}$ & 3.65 &0.18 \\
  2MASX J16531506+2349431 & $1.1\times10^{43}$ & $4.9\times10^{45}$  & $1.5\times10^{55}$ & 4.40 & 0.35\\
  2MASX J14054117+4026326 & $4.6\times10^{42}$ &$2.1\times10^{45}$  & $6.2\times10^{54}$ & 4.00 &0.26\\
  2MASX J13003807+5454367 & $4.8\times10^{42}$ & $2.2\times10^{45}$  & $6.5\times10^{54}$ & 3.54 & 0.15 \\
  2MASX J11001238+0846157 & $8.5\times10^{42}$ & $3.9\times10^{45}$ & $1.2\times10^{55}$ & 3.60 & 0.17\\
  2MASX J08025293+2552551 & $8.0\times10^{42}$ & $3.6\times10^{45}$  & $1.1\times10^{55}$ & 4.59 &0.39 \\
  2MASX J07594101+5050245 & $8.6\times10^{42}$ & $3.9\times10^{45}$  & $1.2\times10^{55}$ & 4.70 & 0.41\\ 
  FIRST J120041.4+314745 & $1.3\times10^{43}$ & $5.8\times10^{45}$ & $1.7\times10^{55}$ & 3.52 & 0.15 \\
  B2 1435+30 & $7.1\times10^{42}$& $3.2\times10^{45}$ &  $9.6\times10^{54}$ & 4.33 & 0.33 \\

\end{tabular*}
\end{table*}

\section{Results}
\label{sec:sec_4}

\begin{figure*}
  \centering
 
 \begin{minipage}[b]{0.45\textwidth}
  \includegraphics[width=8.65cm]{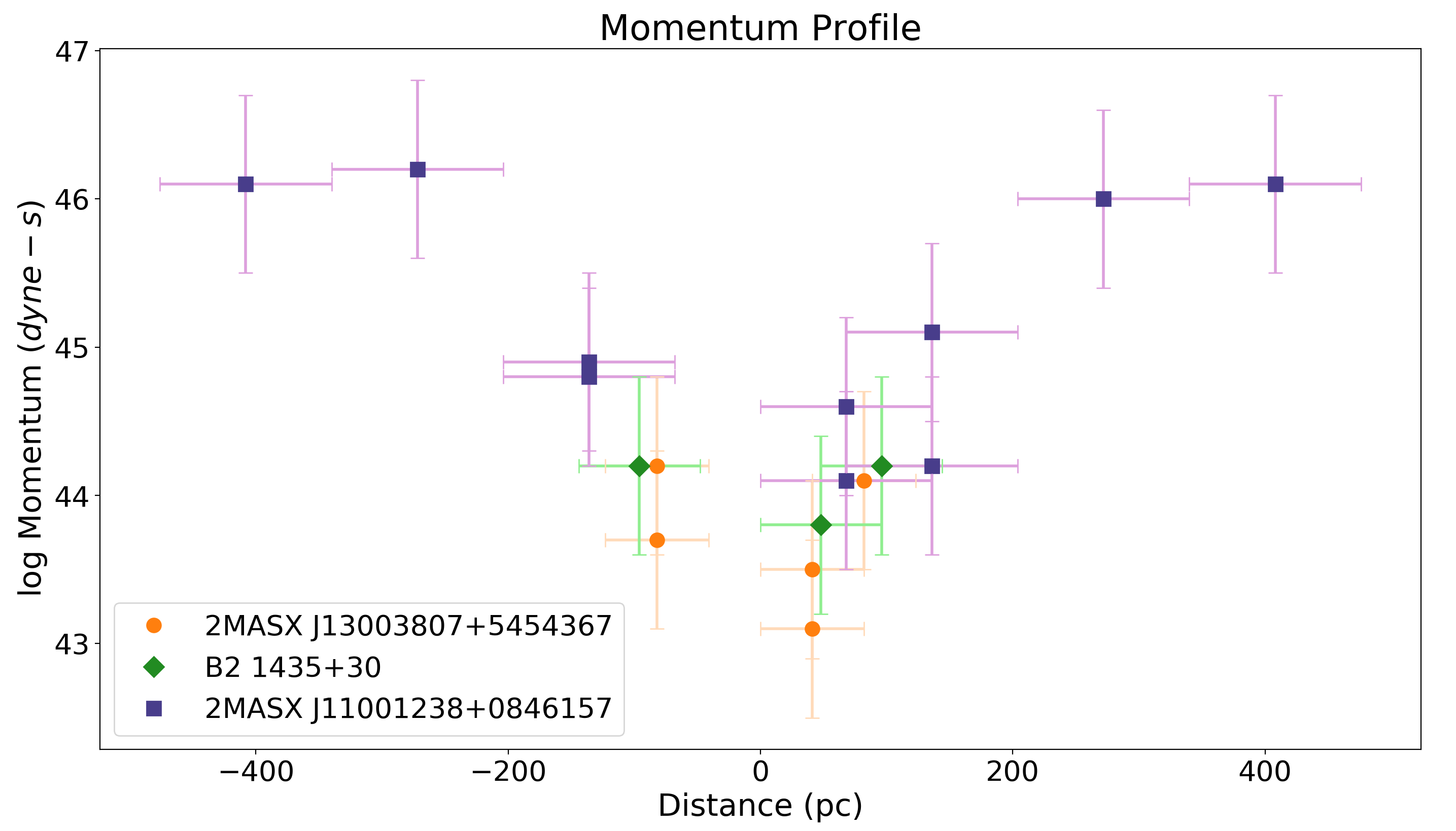}
 \end{minipage}\qquad 
 \begin{minipage}[b]{0.45\textwidth}
  \includegraphics[width=8.65cm]{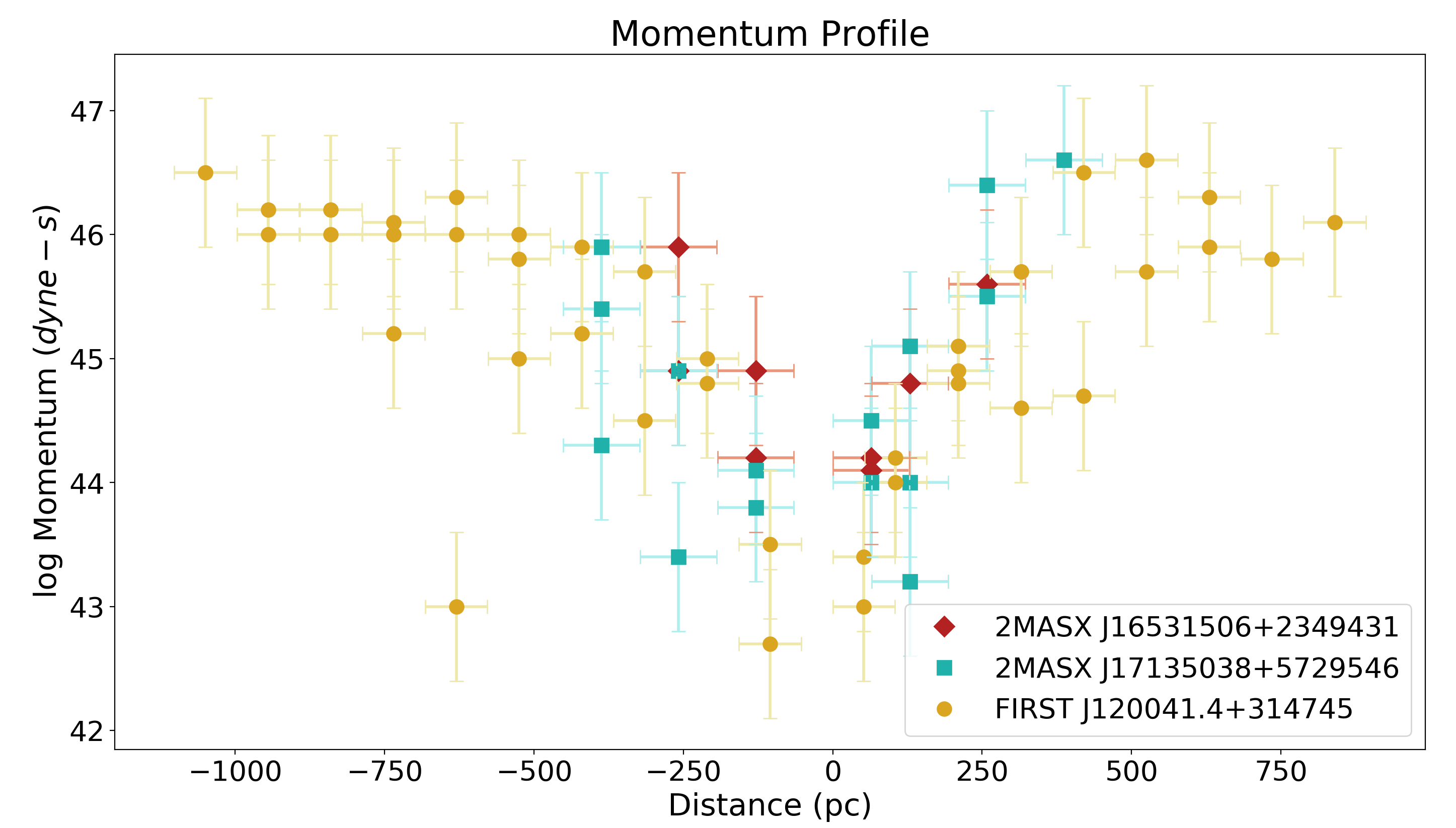}
 \end{minipage}\qquad
 \begin{minipage}[b]{0.45\textwidth}
  \includegraphics[width=8.65cm]{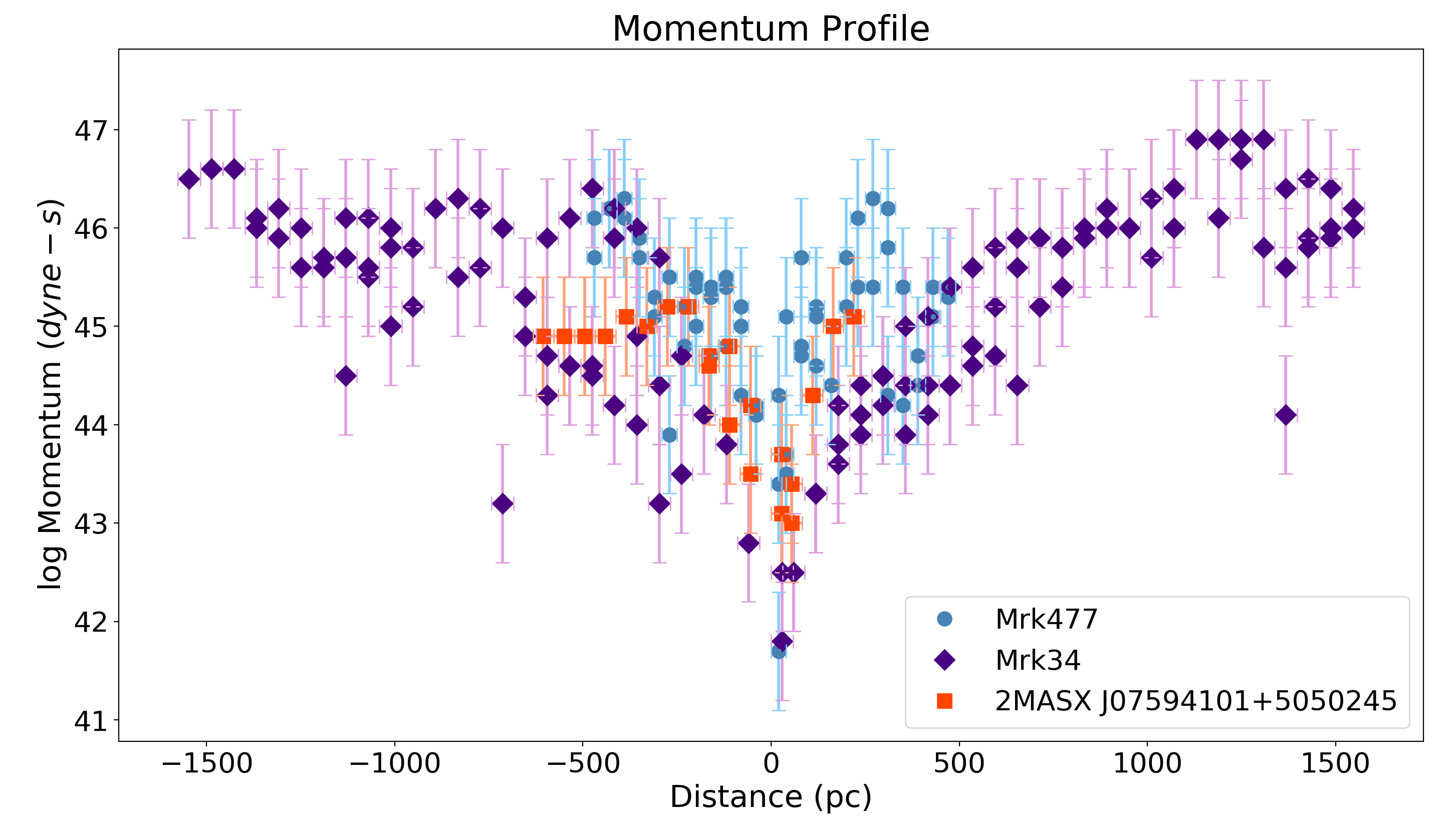}
 \end{minipage}\qquad 
 \begin{minipage}[b]{0.45\textwidth}
  \includegraphics[width=8.65cm]{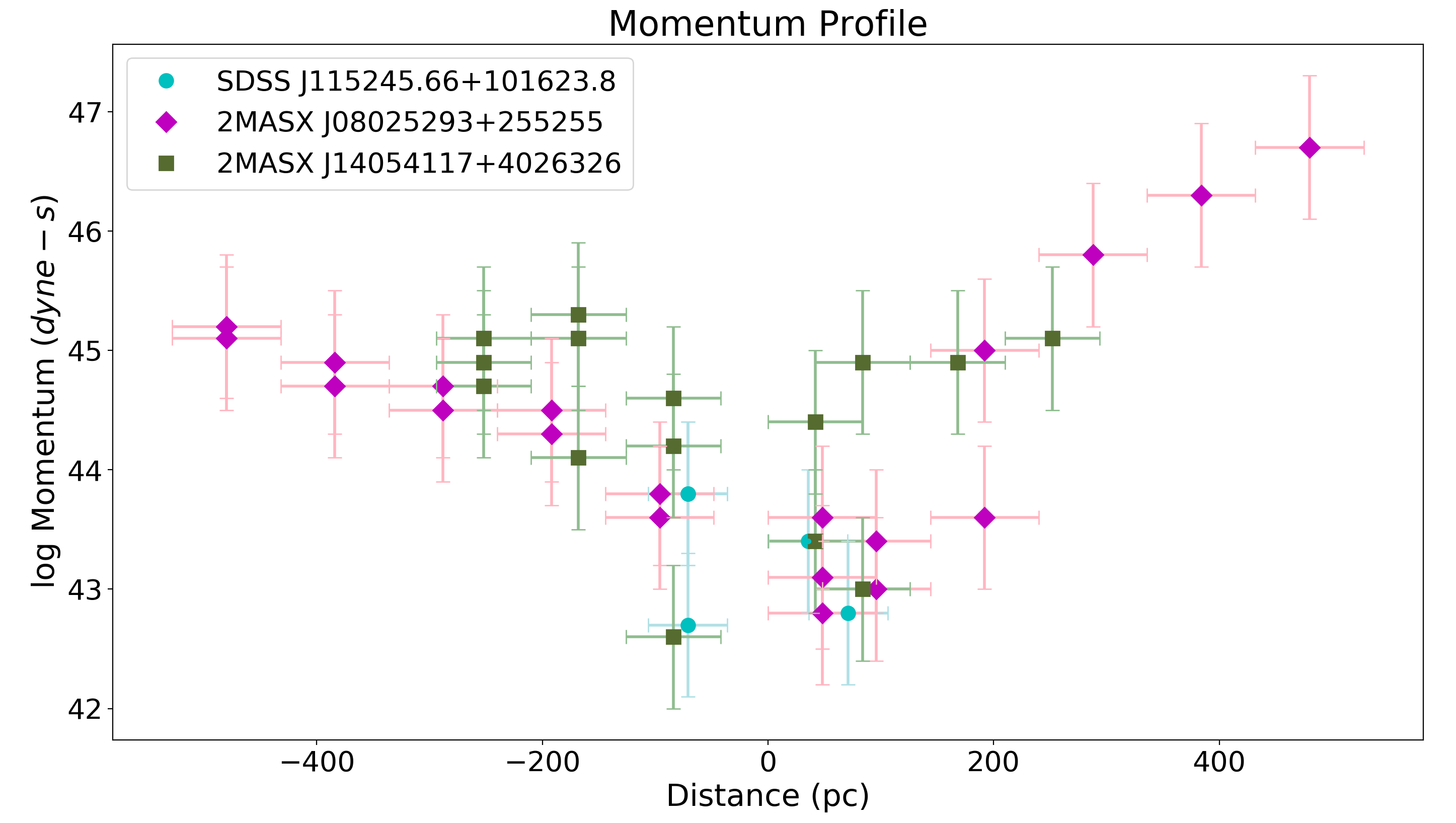}
 \end{minipage}\qquad

\caption{ Momentum Profiles for all targets. See section \ref{sec:sec_4}}
\label{fig:Fig_8}
 \end{figure*}

\begin{figure*}
  \centering

 \begin{minipage}[b]{0.45\textwidth}
  \includegraphics[width=8.65cm]{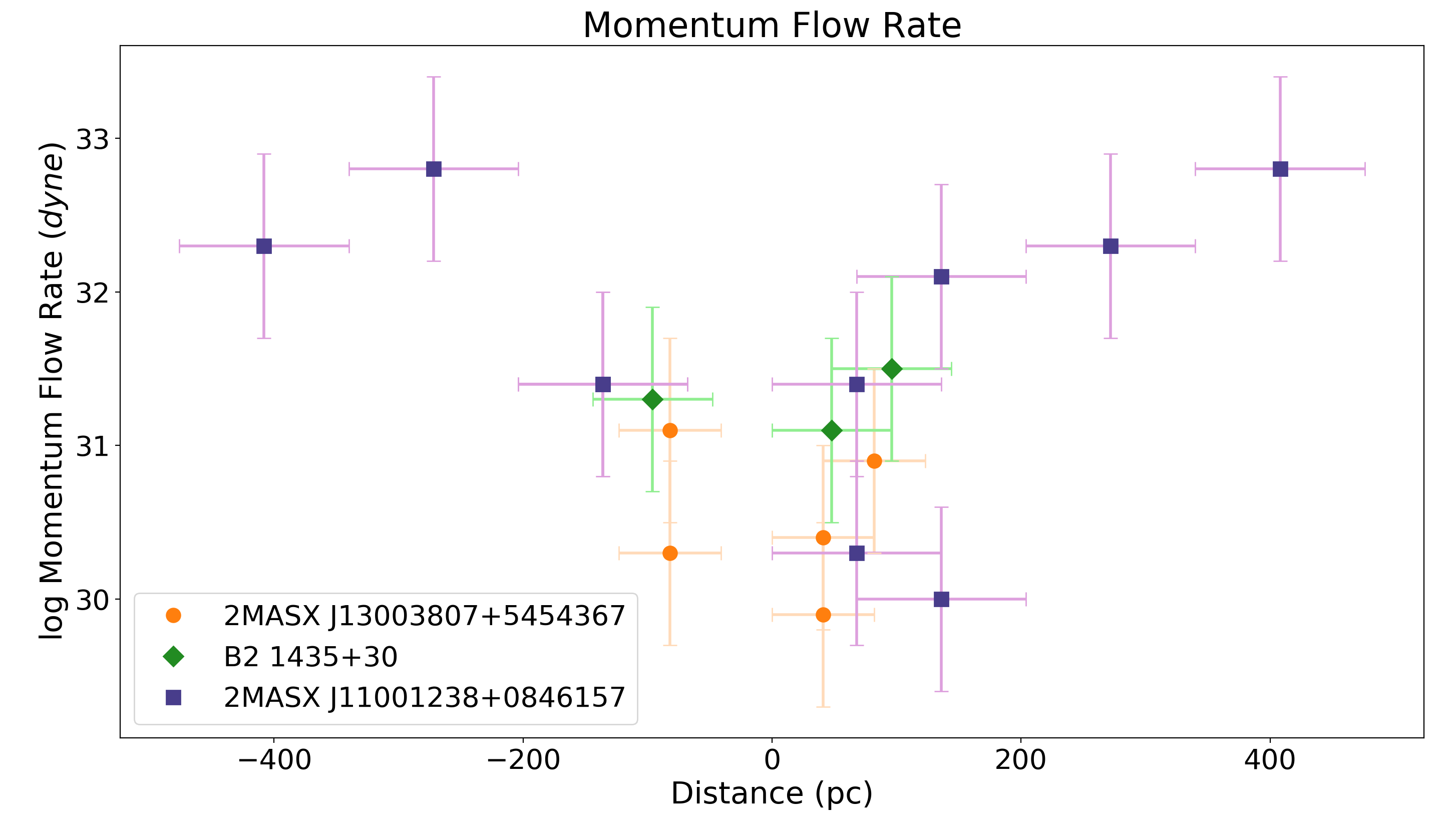}
 \end{minipage}\qquad 
 \begin{minipage}[b]{0.45\textwidth}
  \includegraphics[width=8.65cm]{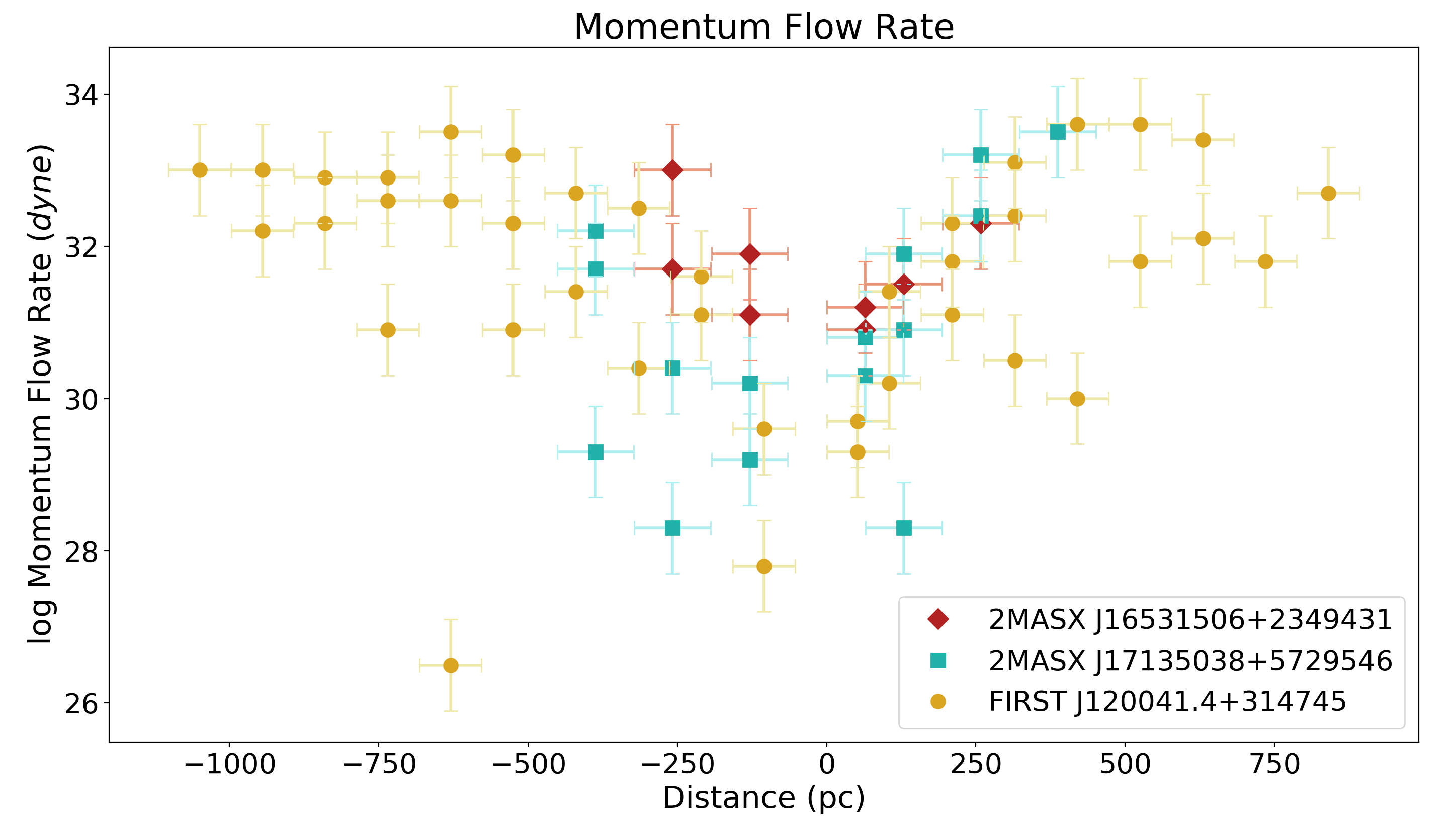}
 \end{minipage}\qquad 
 \begin{minipage}[b]{0.45\textwidth}
  \includegraphics[width=8.65cm]{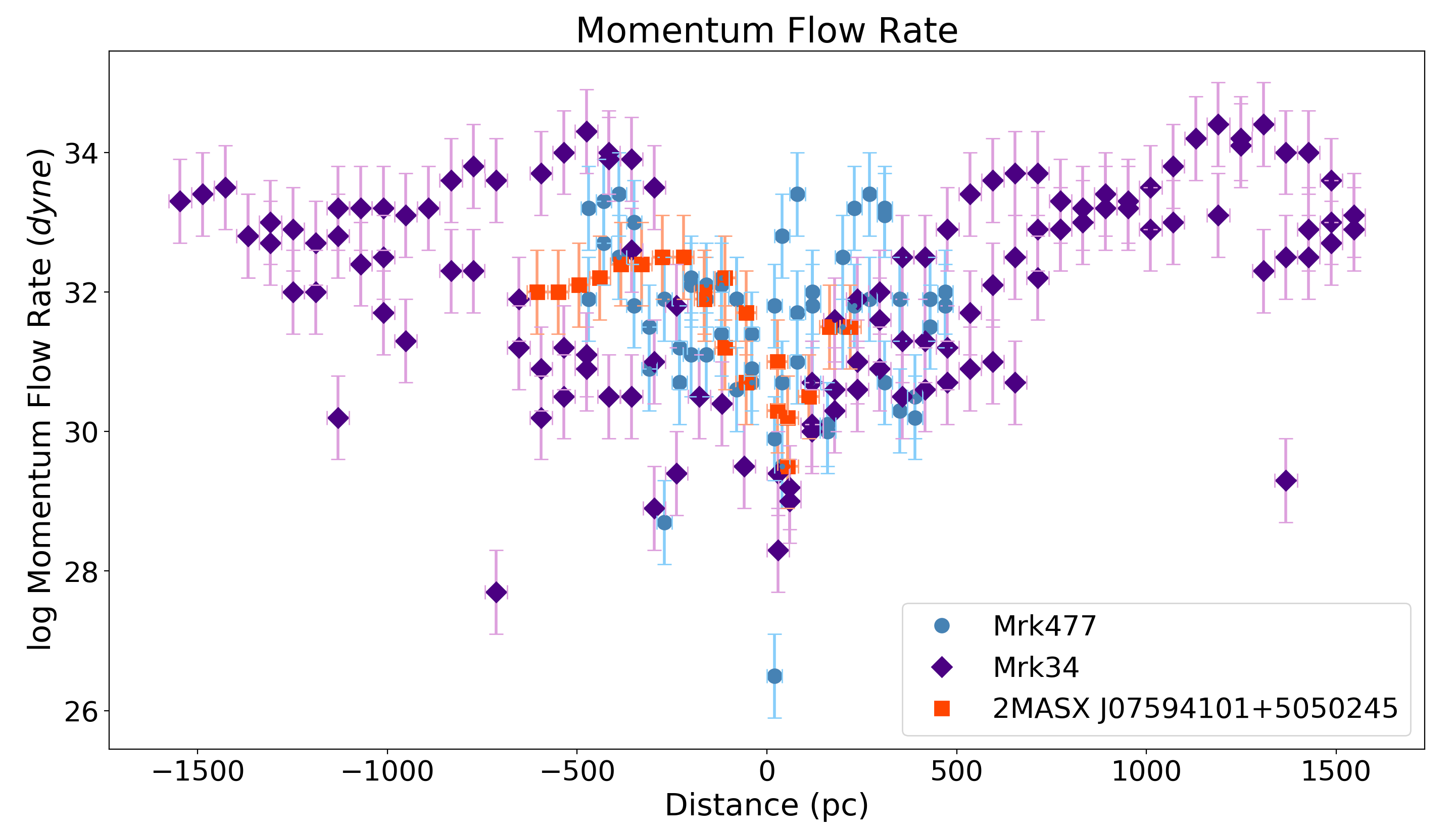}
 \end{minipage}\qquad
 \begin{minipage}[b]{0.45\textwidth}
  \includegraphics[width=8.65cm]{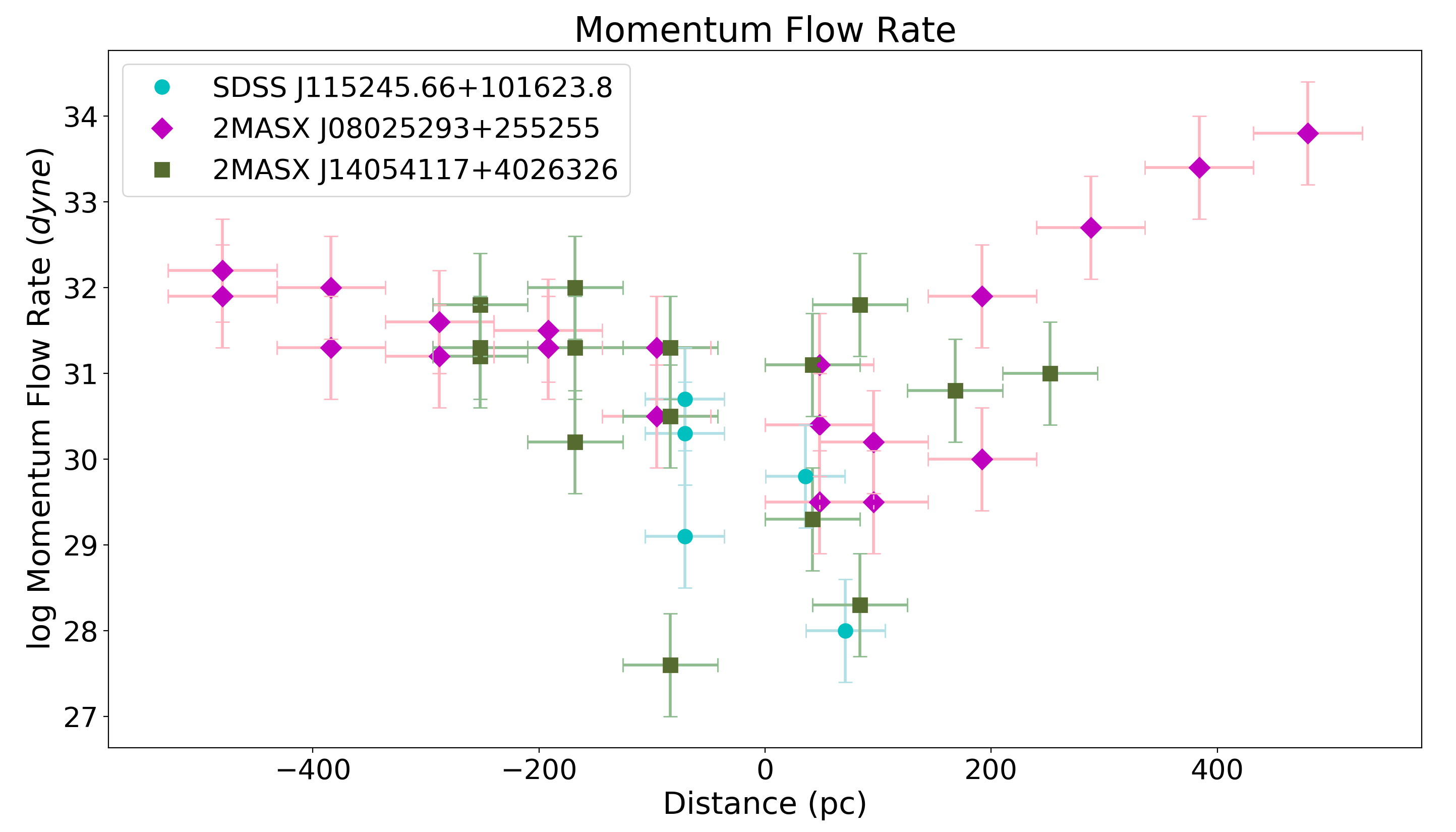}
 \end{minipage}\qquad

\caption{Momentum flow rates. See section \ref{sec:sec_4}}
\label{fig:Fig_9}
\end{figure*}

We present our mass profiles, outflow rates and other kinematic properties as functions of distance from the SMBH in Figures \ref{fig:Fig_4} through Figure \ref{fig:Fig_9}. The quantities shown are the values within each bin of length $\delta r$. Table \ref{tab:Table_3} gives the total masses outflowing and the maxima of the kinematic properties. Table \ref{tab:Table_4} gives the radii of the peaks in these quantities.\par

  Among the QSO2s in our sample, the outflow has a maximum radial extent that extends from 130 pc to 1,600 pc from the nucleus and contains a total ionised gas mass ranging from $4.6^{+13.8}_{-3.45}\times 10^{3}$ ${\rm M\textsubscript{\(\odot\)}}$ to  $3.4^{+10.2}_{-2.55}\times 10^{7}$ ${\rm M\textsubscript{\(\odot\)}}$, in the outflow region (Figure \ref{fig:Fig_4}). The kinematics at further distances are consistent with disturbance and rotation, but not with outflows, as discussed in \citetalias{F2018}.\par 
 For all targets in our sample, the total ionised mass at the innermost points is low (see Figure \ref{fig:Fig_4}), which can be explained by the fact that the dense gas in this region radiates more efficiently (see \citealt{OF}), leading to smaller masses being required to produce the same observed [O~III] emission. Additionally, the area sampled in the annuli is an increasing function of radius, meaning that less mass is added from the images at smaller distances.
 
The peak in the mass outflow rate peaks from $9.3^{+27.9}_{-7.0}\times10^{-3}$ ${\rm M\textsubscript{\(\odot\)}}$  ${\rm yr^{-1}}$, for B2 1435+30, to $10.3^{+30.9}_{-7.7}$ ${\rm M\textsubscript{\(\odot\)}}$ ${\rm yr^{-1}}$, for Mrk 34 (Figure \ref{fig:Fig_5}), at a distance varying from 100 pc to 1,260 pc from the nucleus (see Table \ref{tab:Table_4}) and then decrease at larger distances.\par 
The maximum kinetic energy for the targets extends from $8.2^{+24.6}_{-6.1}\times10^{50}$ ergs, for SDSS J115245.66+101623.8, to  $2.6^{+7.8}_{-1.9}\times10^{54}$ ergs, for Mrk 34. The total kinetic luminosity ranges from $3.5^{+10.5}_{-2.6}\times 10^{-8}$ of $L_{bol}$, for SDSS J115245.66+101623.8, to $4.1^{+12.3}_{-3.1}\times 10^{-3}$ of $L_{bol}$, for Mrk 34. \par

The radiation pressure force from the bolometric luminosity, $\frac{L_{bol}}{c}$, ranges from $5.4^{+16.2}_{-4.0}\times 10^{34}$ dyne, for SDSS J115245.66+101623.8, to $1.9^{+5.7}_{-1.4}\times 10^{35}$ dyne, for FIRST J120041.4+314745. The peak momentum flow rate extends from $7.5^{+22.5}_{-5.6}\times10^{30}$ dyne, for SDSS J115245.66+101623.8, to $2.8^{+8.4}_{-2.1}\times10^{34}$ dyne, for Mrk 34. Thus, considering the maximum momentum flow rate among our sample, the peak outflow momentum rate is $\sim$ 30\% of the AGN’s radiation pressure force, although it is a much smaller percentage for most of our sample.\par

\begin{figure*}
  \centering
   \begin{minipage}[b]{0.45\textwidth}
  \includegraphics[width=8.65cm]{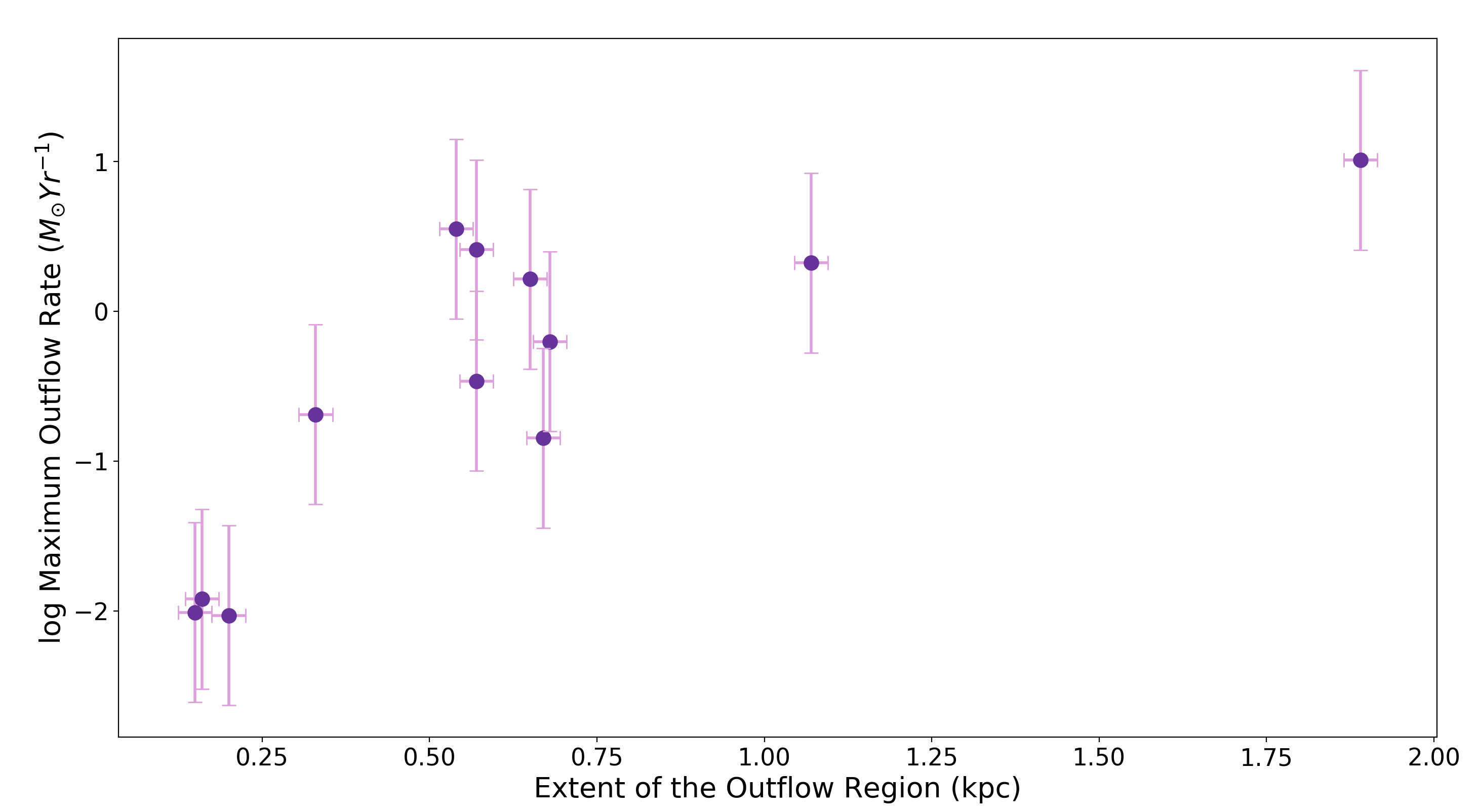}
 \end{minipage}\qquad 
 \begin{minipage}[b]{0.45\textwidth}
  \includegraphics[width=8.65cm]{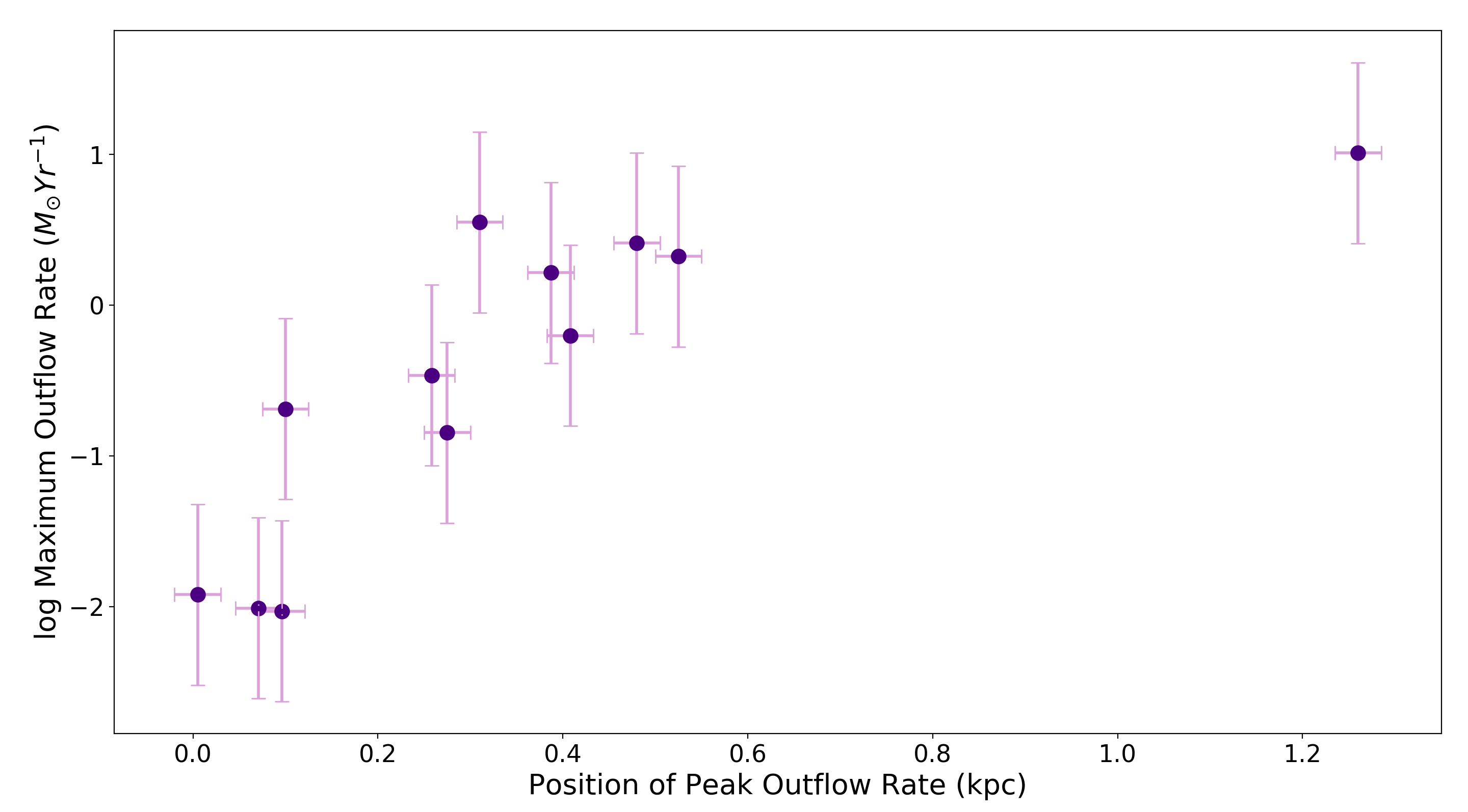}
 \end{minipage}\qquad 
\caption{For a majority of our sample, it is possible to see a relation between the maximum outflow rate and $R_{out}$ (left panel) and the position where the peak outflow rate occurs and the maximum outflow rate for each target (right panel).}
\label{fig:Fig_10}
\end{figure*}

The quantities displayed are the value contained within each bin of width $\delta r$. In addition, we neglect contributions to the mass outflow rates and energetics from the significant FWHM of the emission lines, which could be due, for example, to the ablation of gas off the spiral dust lanes.\par

\section{Discussion}
\label{sec:sec_5}

The maximum kinetic luminosities of the outflow for our sample are presented in Table \ref{tab:Table_3} and are $(3.4^{+10.2}_{-2.5}\times 10^{-8}  - 4.9^{+14.7}_{-3.7}\times 10^{-4}$ of the AGN bolometric luminosities. This does not approach the $5.0\times 10^{-3}$ - $5.0\times 10^{-2}$ range used in models of efficient feedback \citep{DM2005, HE2010}. The low ratios of  $\dot E$/$L_{bol}$ were also found in other samples by \citet{BN2019} and \citet{D2020}, however, these are much less than those derived from relativistic outflows (UFOs) observed in the X-ray spectra of some AGN \citep{Bi2019}. \par
We also study the relation between the maximum outflow rate for each target and the position where the maximum outflow rate occurs (Figure \ref{fig:Fig_10}, right panel), as well as the relation between the maximum outflow rate and $R_{out}$ for each target in our sample (Figure \ref{fig:Fig_10}, left panel). For a majority of the targets, the maximum outflow rate appears to be correlated with $R_{out}$, since the further away the gas is from the SMBH, the larger the amount of gas is needed to produce the same amount of ionised gas mass, as the density drops. \par 
After concluding that these QSO2s do not produce effective feedback, based on models previously discussed, we calculate how much mass an outflow would have to have to produce a $\dot E$ = 0.5\% of $L_{bol}$, which is considered to be the value where feedback effects could be relevant\footnote[7]{The maximum $\dot E$ does not necessarily occur at the point where the target has its maximum outflow rate, since $\dot E \sim  v_{out}^3$.} \citep{HE2010}.

\begin{figure*} 
 \begin{minipage}[b]{0.45\textwidth}
  \includegraphics[width=8.65cm]{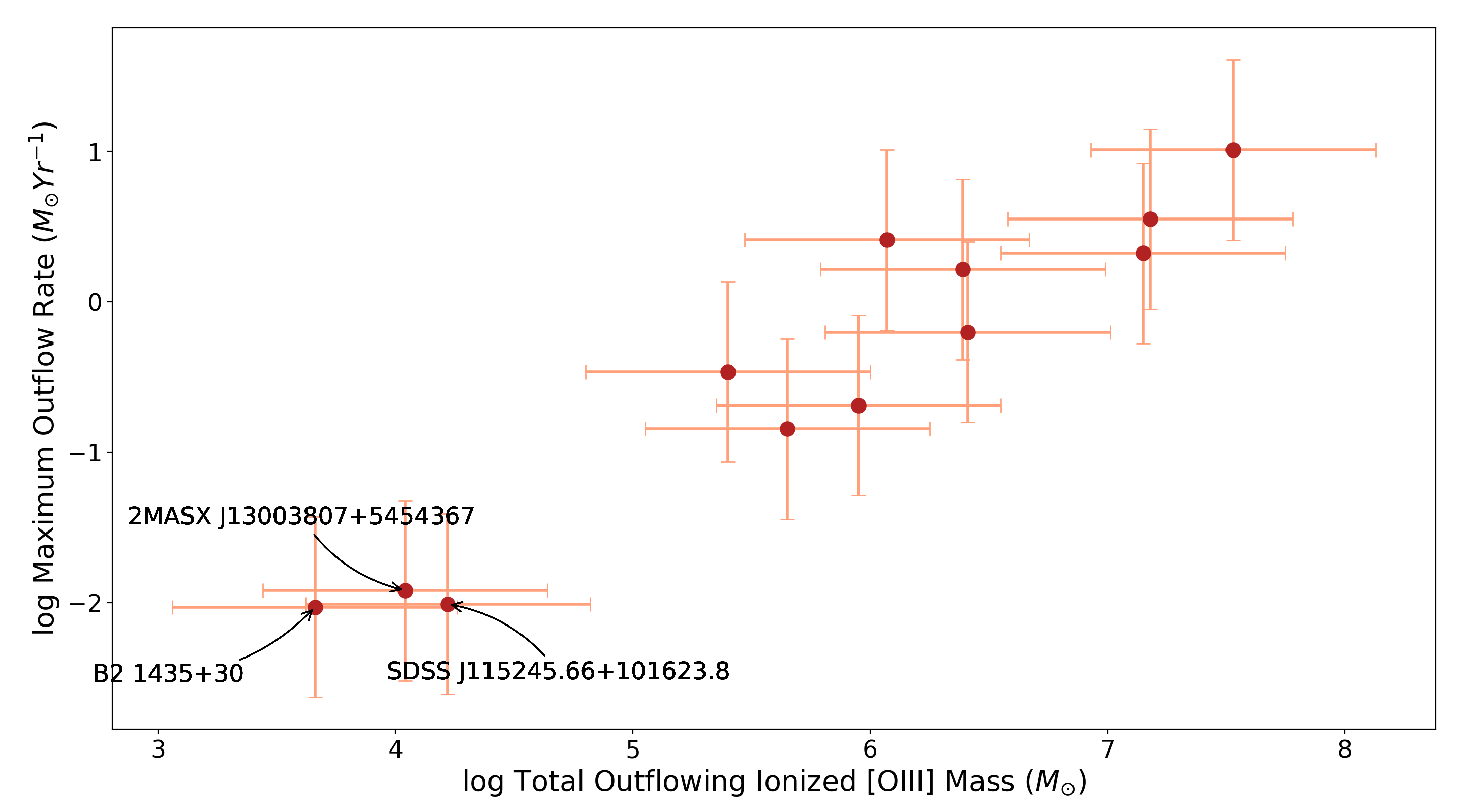}
 \end{minipage}\qquad 
  \begin{minipage}[b]{0.45\textwidth}
  \includegraphics[width=8.65cm]{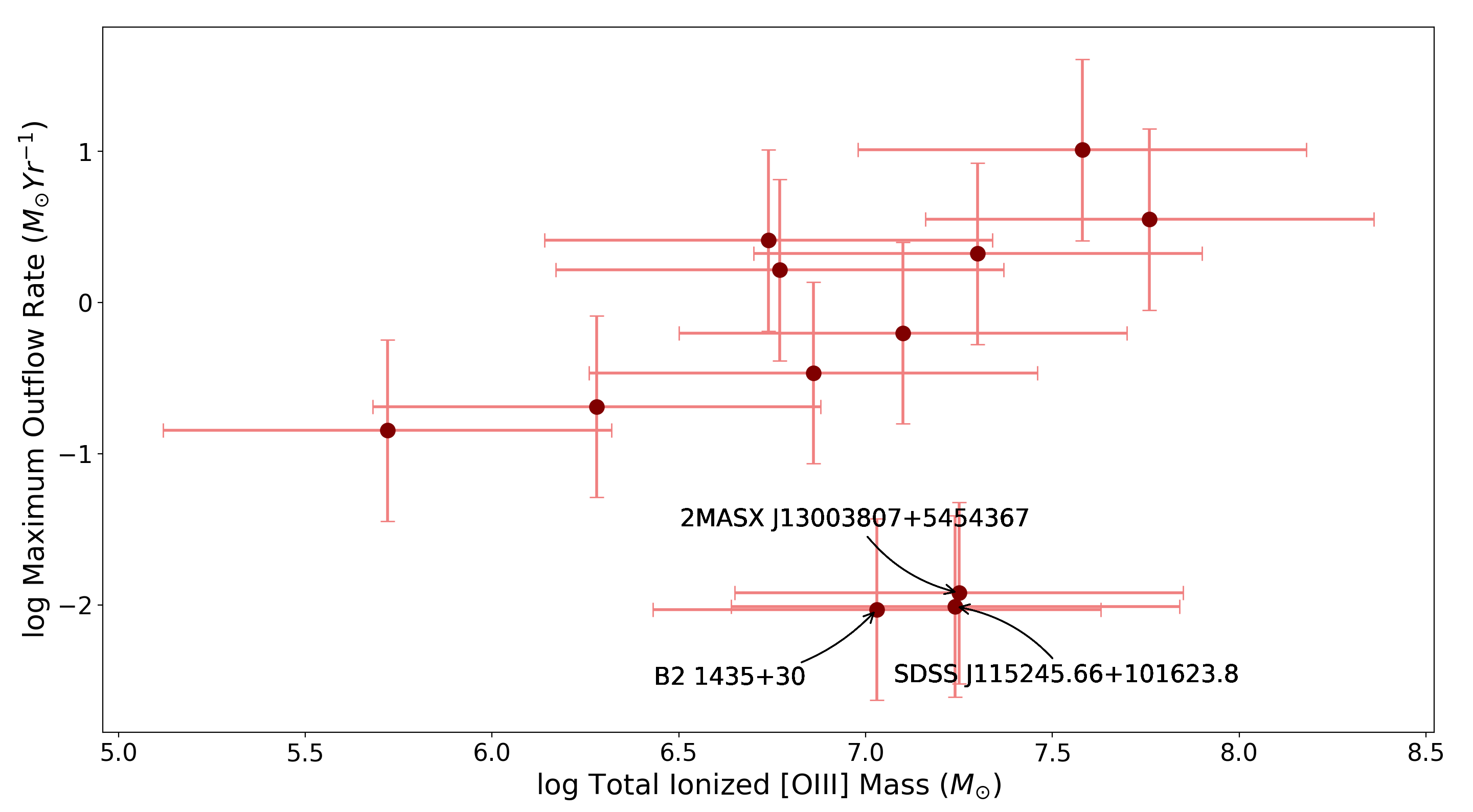}
 \end{minipage}\qquad 
\caption{\label{fig:Fig_11}The left panel shows the relation between the maximum outflow rate for each target and the total ionised mass in the outflow range. Right panel shows the relation between the maximum outflow rate for each target and the total ionised mass including the regions outside the outflow range.}
\end{figure*}

We calculate the required mass for Mrk 34, at its point of maximum $\dot E$, using the highest deprojected outflow velocity component calculated using the fitting routine by \citet{F2017} for that same position, $v_{out}=1.4\times10^3$ km/s. We choose Mrk 34 since, among the targets in our sample, it is the target that shows the highest outflow rate and most extended outflows. We find that, in order to reach the lower limit of $\dot E$/$L_{bol}$ = 0.5\%, Mrk 34 would have to possess an outflow rate of $20$ ${\rm M\textsubscript{\(\odot\)}}$  ${\rm yr^{-1}}$, in contrast to our measured value of $1.97^{+5.9}_{-1.5}$ ${\rm M\textsubscript{\(\odot\)}}$  ${\rm yr^{-1}}$. This corresponds to a mass, at that position, of $M_{0.5\%}=8.2\times10^{5}$ ${\rm M\textsubscript{\(\odot\)}}$, as opposed to our measured value of $8.1^{+24.3}_{-6.1}\times10^{4}$ ${\rm M\textsubscript{\(\odot\)}}$. \par 
We also calculate the maximum amount of mass, $M_{max}$, that Mrk 34 could have at this distance, i.e., the maximum amount of gas that could be at this distance if the entire solid angle around the SMBH were covered. Specifically:

\begin{equation}
    M_{max}(r)= {4\pi r^2 (\mu  m_{p} N_{H})}
\end{equation}

\noindent where $r$ is the deprojected radial distance. We are assuming the gas is distributed in a complete shell at this distance. 
We find that the maximum amount of mass that Mrk 34 could possess at this position is $M_{max}(r)=1.3\times 10^8$ ${\rm M\textsubscript{\(\odot\)}}$, which is greater than the value for $M_{0.5\%}$. However, only a small fraction of that mass is emitting [O~III] radiation.\par 
If we consider the position where Mrk 34 has its peak outflow rate, 1.26 kpc, in order to calculate how much mass it would need to reach 0.5\% of its bolometric luminosity, and using the highest deprojected outflow velocity at that position, $v_{out}=6.1\times10^2$ km/s, we find that there would need to be an outflow rate of $109$ ${\rm M\textsubscript{\(\odot\)}}$  ${\rm yr^{-1}}$, in contrast to our measured value of $4.1^{+12.3}_{-3.1}$ ${\rm M\textsubscript{\(\odot\)}}$  ${\rm yr^{-1}}$ (in \citealt{R2018} the peak outflow rate occurs at $\sim$ 0.5 kpc, where the deprojected outflow velocity is $\sim$ 2,000 km/s, which would require an outflow rate of $\sim$ 40 ${\rm M\textsubscript{\(\odot\)}}$  ${\rm yr^{-1}}$). This corresponds to a mass, at that position, of $1.0\times10^7$ ${\rm M\textsubscript{\(\odot\)}}$, as opposed to our measured value of $4.0^{+12.0}_{-3.0}\times10^5$ ${\rm M\textsubscript{\(\odot\)}}$. We find that the maximum amount of mass that Mrk 34 could possess at this position is $M_{max}(r)=7.4\times 10^8$ ${\rm M\textsubscript{\(\odot\)}}$, which is also greater than the value for $M_{0.5\%}$.\par 

\begin{table*}

\caption{Numerical results for the mass and energetic quantities for the outflowing gas component in each target of our sample. All results account for the gas within the outflow region. Column 2 is the gas mass in units of
$10^{5} M\textsubscript{\(\odot\)}$, Column 3 is  peak mass outflow rate within the outflow region, Column 4 is the peak kinetic energy, Column 5 is  the peak kinetic luminosity, Column 6 is the peak momentum, and Column 7 is the peak momentum flow rate for each QSO2. All listed values have a factor of $\sim$ 4 uncertainty, as discussed in section \ref{sec:sec_2.4}.}
\label{tab:Table_3}

\smallskip
\begin{tabular*}{\textwidth}{@{\extracolsep{\fill}}lcccccc}

  Target & Total Mass & Maximum $\dot M $ & Maximum E  & Maximum $\dot E$ & Maximum p & Maximum $\dot p$ \\
  & ${\rm(10^{5}M\textsubscript{\(\odot\)})}$ & ${\rm(M\textsubscript{\(\odot\)}/yr)}$ & ${\rm(10^{53}ergs)}$ & ${\rm(10^{41} erg/s)}$ & ${\rm(10^{46} dyne-s)}$ & ${\rm(10^{34} dyne)}$\\
  (1)&(2)&(3)&(4)&(5)&(6)&(7)\\

  SDSS J115245.66+101623.8 & 0.17 & 0.01 & $<$0.01 & $<$0.01 & 0.01 & $<$0.01 \\
  MRK 477 & 152.0 & 3.55 & 1.34 & 0.75 & 2.09 & 0.3 \\
  MRK 34  & 337.0 & 10.3 & 26.20 & 12.80 & 9.15 & 2.83 \\
  2MASX J17135038+5729546 & 24.40 & 1.64 & 6.50 & 0.51 & 4.17 & 0.33 \\
  2MASX J16531506+2349431 & 2.54 & 0.34 & 2.23 & 0.25 & 0.95 & 0.10 \\
  2MASX J14054117+4026326 & 8.94 & 0.20 & 0.14 & $<$0.01 & 0.34 & 0.01 \\
  2MASX J13003807+5454367 & 0.11 & 0.01 & 0.02 & $<$0.01 & 0.02 & $<$0.01 \\
  2MASX J11001238+0846157 & 25.60 & 0.63 & 1.32 & 0.07 & 1.68 & 0.06 \\
  2MASX J08025293+2552551 & 11.80 & 2.58 & 9.54 & 1.27 & 4.86 & 0.64 \\
  2MASX J07594101+5050245 & 4.44 & 0.14 & 0.29 & 0.07 & 0.15 & 0.03 \\ 
  FIRST J120041.4+314745 & 140.0 & 2.11 & 6.34 & 1.00 & 4.35 & 0.39 \\ 
  B2 1435+30 & 0.04 & $<$0.01 & 0.45 & $<$0.01 & 0.02 & $<$0.01 \\

\end{tabular*}
\end{table*}

These results tell us that the required amount of gas for efficient feedback is less than what we would estimate for a covering factor of unity, which means that it is theoretically possible to have an $\dot E$ = 0.5\% of $L_{bol}$. One explanation for the low values we are obtaining for the mass outflow rate is that the source of the outflow, e.g., cold molecular and possibly atomic gas in the disk \citep{F2017}, has a very low covering factor compared to a sphere at its location. \par 
 Another possibility is that the gas does not remain in the state in which it emits [O~III] for long. If it is not confined by an external medium, e.g., a lane of gas and dust in the host disk, it will rapidly thermally expand. As it does so, the density drops, and the ionisation state of the gas increases to the point where it becomes X-ray emission line gas (as suggested by \citealt{K2020}). In this case, the outflows could be dominated by the X-ray emitting gas. We will be exploring this scenario in a future paper (Trindade Falcão et al. in preparation). \par 
Another characteristic of these targets is the relation between the total ionised mass and the maximum outflow rate for each AGN.\par 
We plot these quantities in two different ways: \par 
\noindent 1) considering the total ionised mass only within the outflow range (Figure \ref{fig:Fig_11}, left panel);\par 
\noindent 2) considering the ionised mass including the mass outside the outflow range (Figure \ref{fig:Fig_11}, right panel). \par 
When comparing the two plots, we find three AGNs, SDSS J115245.66+101623.8, 2MASX J13003807+5454367 and B2 1435+30, are separate from the remainder of the QSO2 sample. Their shift in positions from the left panel to the right panel on Figure \ref{fig:Fig_11} is due to their very extended [O~III] emission, based on the ACS images, despite having very weak outflow rates.\par 

\begin{table*}

\caption{Deprojected distance from the nucleus for peak measurements of mass outflow rates (Column 2), kinetic luminosities (Column 3) and momentum flow rates (Column 4). Column 5 lists the highest deprojected outflow velocity at the maximum outflow rate position.}
\label{tab:Table_4}

\smallskip
\begin{tabular*}{\textwidth}{@{\extracolsep{\fill}}lcccc}

 Target & Position of Peak $\dot M$  & Position of Peak $\dot E$  & Position of Peak $\dot p$ & $v_{out}$ at peak $\dot M$ \\
  & (pc) & (pc) & (pc) & (km/s)\\
 (1)&(2)&(3)&(4)&(5)\\

  SDSS J115245.66+101623.8 & 70 & 70 & 70 & 70.9 \\
  MRK 477 & 310 & 80 & 310 & 22.9  \\
  MRK 34  & 1260 & 480 & 1320 & 341   \\
  2MASX J17135038+5729546 & 390 & 390 & 390 & 312  \\
  2MASX J16531506+2349431 & 260 & 260 & 260 & 498  \\
  2MASX J14054117+4026326 & 170 & 170 & 170 & 42.4  \\
  2MASX J13003807+5454367 & 50 & 50 & 50 & 206   \\
  2MASX J11001238+0846157 & 270 & 410 & 410 & 144 \\
  2MASX J08025293+2552551 & 480 & 480 & 480 & 393 \\
  2MASX J07594101+5050245 & 275 & 275 & 275 & 379 \\ 
  FIRST J120041.4+314745 & 525 & 840 & 525 & 325 \\ 
  B2 1435+30 & 100 & 100 & 100 & 518  \\

\end{tabular*}
\end{table*}

To better understand the conditions under which the gas can be efficiently accelerated we can use the velocity calculation for radiatively driven outflows, discussed by \citealt{D2007}, where: 

\begin{equation}
    v(r) = \sqrt{\int_{r_1}^{r} {A_{1}L_{bol}\frac{\mathcal{M}}{r^2}-A_{2}\frac{M(r)}{r^2}}\, dr}
\end{equation}
\noindent where $\mathcal{M}$ is the Force Multiplier, i.e., the ratio of the total absorption cross section to the Thomson cross section, and $M(r)$ is the enclosed mass at the distance $r$, determined from the radial mass distribution of the host galaxy, including the bulge. The constants $A_{1}$ and $A_{2}$ are given in \citealt{D2007}.\par 

Taking into account the relevant parameters in Equation 13, specifically, $L_{bol}$ and $M(r)$, there are two possibilities that can explain the distinct characteristics of these three QSO2s:\par 
1) The AGN was in a low state until recently. We can rule out the possibility that it was "off" completely by estimating the recombination times for the [O~III] gas, as follows:

\begin{equation}
    \tau = \frac{1}{n_{H} \alpha_{rec}}
\end{equation}

The total recombination rate for ${\rm O^{++}}$  -  ${\rm O^{+}}$ is $\alpha_{rec}$ = ${\rm 2.52 \times 10^{-12}}$ ${\rm cm^3}$ ${\rm sec^{-1}}$, at $10^{4}K$ \citep{Nh1999}. Based on our model densities, the recombination times (Equation 14) are relatively short (the minimum recombination times are $\sim$ 3 - 4 years and the maximum range from 245 years to 370 years), therefore, it is possible that the AGN could have turned completely off. However, even if the AGN was on, it appears to be in too weak of a state to accelerate the [O~III] gas at these distances. \par 
If the AGN was in a weak state, it could have enough ionising radiation to produce [O~III]. However, the $L_{bol}$ would have been too low, up to recently, to accelerate the gas, that it would not be able to accelerate the gas, since the first term in Equation 13 becomes smaller compared to the gravitational deceleration term. \par 

If we assume that the existence of outflows close to the AGN mean that the AGN is back to a high state, we can estimate how long ago it entered into this high state by using the size of the outflow regions and calculating the light crossing time for each target. Our calculations show that for SDSS J115245.66+101623.8, 2MASX J13003807+5454367 and B2 1435+30 these values are $\sim$ 230, 270 and 310 years ago, respectively.\par 

2) The AGN has not varied in luminosity, but the three targets have different mass distributions than the other QSO2s in our sample. This possibility can be considered as two distinct cases:\par 
a) The bulge mass is more centrally peaked. If this is the case, the second term in Equation 13 starts to dominate closer to the nucleus. However, the velocities and outflow rates for these three targets are similar to those of the rest of the sample in the same distance range (see Table \ref{tab:Table_4}). This suggests that this scenario is unlikely. \par 
b) These objects possess a mass component which starts to dominate some distance from the AGN, as in Mrk 573 \citep[see][figure 14]{F2017}. In this case, outflows can be generated close to the AGN, but, when this outer component starts to dominate, the gas cannot be accelerated. However, in order to explore this possibility, we would need deeper (higher S/N) continuum images to derive the stellar mass profiles. \par

We are currently studying the dynamics of  the outflows in these QSO2s, which will address this issue in more detail (Trindade Falcão et al. in preparation).  \par

\section{Conclusions}
\label{sec:sec_6}

We use long-slit spectroscopy, [O~III] imaging, and Cloudy photoionisation models to determine the mass outflow rates and energetics as functions of distance from the nucleus in a sample of twelve nearby (z $<$ 0.12) luminous ($L_{bol} > 1.6 \times 10^{45}$ ${\rm erg}$ ${\rm s^{-1}}$) QSO2s. Our results are as follows:\par
 1. The outflows contain a total ionised gas mass ranging from $4.6^{+13.8}_{-3.4}\times 10^{3}{\rm M\textsubscript{\(\odot\)}}$, for B2 1435+30, to  $3.4^{+10.2}_{-2.5}\times 10^{7}{\rm M\textsubscript{\(\odot\)}}$, for Mrk 34, with a total kinetic energy varying between $8.9^{+26.7}_{-6.7}\times 10^{50}$ ${\rm ergs}$, for SDSS J115245.66+101623.8, and $2.9^{+9.0}_{-2.2}\times 10^{55}$ ${\rm ergs}$, for Mrk 34. \par
 
 2. \citetalias{F2018} found that the outflows extend to a maximum of 1,600 pc. Our results show that these outflows reach a peak outflow rate ranging from $\dot M_{out}(r)=9.3^{+27.9}_{-7.0}\times10^{-3}$ ${\rm M\textsubscript{\(\odot\)}}$ ${\rm yr^{-1}}$, for B2 1435+30, to $10.3^{+30.9}_{-7.7}$ ${\rm M\textsubscript{\(\odot\)}}$ ${\rm yr^{-1}}$, for Mrk 34, at distances between 100 pc and 1260 pc from the SMBH. \par
 
 3. The maximum kinetic luminosity of the outflow ranges from $3.4^{+10.2}_{-2.5}\times 10^{-8}$ of the AGN bolometric luminosity for SDSS J115245.66+101623.8 to $4.9^{+14.7}_{-3.7}\times 10^{-4}$ of the AGN bolometric luminosity for Mrk 34\footnote[8]{Our maximum kinetic luminosity value for Mrk 34 is about a factor of 10 less than that computed by Revalski et al. (2018). The discrepancy is due to the presence of a high mass/low density component, which was not included in our analysis.}. The large range in kinetic luminosity compared to the narrow range in $L_{bol}$ (see Table \ref{tab:Table_2}) is in contrast to the correlation between the kinetic luminosity and $L_{bol}$ suggested by \citet{Fi2017}. Our results indicate that the [O~III] winds are not an efficient feedback mechanism, based on the criteria of \citet{DM2005} and \citet{HE2010}. This means that, not only do the outflows not extend far enough to clear the bulge of gas (\citetalias{F2018}), they also lack the power to do so. \par

 4. As noted above, Mrk 34 is the target in our sample that shows the highest outflow rate and most extended outflows. We calculate what the outflow rates would have had to have been in order to reach the minimum value required for efficient feedback. We find that Mrk 34 would have to have, at its position of maximum $\dot E$, an outflow rate of $20$ ${\rm M\textsubscript{\(\odot\)}}$ ${\rm yr^{-1}}$, corresponding to a mass of $8.2\times 10^5$ ${\rm M\textsubscript{\(\odot\)}}$ at that position. This value is 10 times greater than our measured outflow rate at that same position. We also find that the required mass for efficient feedback is $\sim$ 0.01 times the amount for a covering factor of unity.\par 
 These calculations show that this object could potentially, at these velocities, make efficient outflows, but such energetic outflows are not detected. One possibility is that the lack of [O~III] gas is the result of rapid thermal expansion, with the result that its ionisation state increases to the point where it becomes X-ray emitting gas. In fact, \textit{Chandra} imaging of Mrk 34 has revealed X-ray emission line gas extending the size of the [O~III] emission line region (Fischer et al. in preparation). We are currently exploring this possibility (Trindade Falcão et al. in preparation). Also, neutral and molecular gas could be contributing to the outflow (see section \ref{sec:sec_2.5}) and, altogether, could provide a much higher mass outflow rate \citep{T2015, Bi2019}.\par 
 
5. Three of the targets in our sample show very extended [O~III] emission, but weak outflow rates. Based on their compact outflow regions, but extended [O~III] emission, we study two different scenarios: these AGNs were in a very low state until recently, but have entered a high state, during which they are able to accelerate outflows. Or these AGNs could be housed in more massive host galaxies, prohibiting successful radiative driving at distances greater than a few hundred parsecs.

Based on these results, we do not see the outflows traced by [O~III]-emission-line gas being powerful enough to generate efficient feedback. However, the presence of disturbed gas at larger radial distances (\citetalias{F2018}), suggests that the AGN have an effect outside the outflow regions. One possibility is that this is the result of X-ray winds, which may form from thermal expansion of the [O~III] gas \citep{F2019, K2020}. We are currently exploring this scenario (Trindade Falcão et al., in preparation).

\section*{Acknowledgements}

The authors thank the anonymous referee for helpful comments that improved the clarity of this paper.Support for this work was provided by NASA through grant number HST-GO-13728.001-A from the Space Telescope Science Institute, which is operated by AURA, Inc., under NASA contract NAS 5-26555. Basic research at the Naval Research Laboratory is funded by 6.1 base funding. T.C.F. was supported by an appointment to the NASA Postdoctoral Program at the NASA Goddard Space Flight Center, administered by the Universities Space Research Association under contract with NASA. T.S.-B. acknowledges support from the Brazilian institutions CNPq (Conselho Nacional de Desenvolvimento Científico e Tecnológico) and FAPERGS (Fundação de Amparo à Pesquisa do Estado do Rio Grande do Sul). L.C.H. was supported by the National Key R\&D Program of China (2016YFA0400702) and the National Science Foundation of China (11473002, 11721303). M.V. gratefully acknowledges financial support from the Danish Council for Independent Research via grant no. DFF
4002-00275 and 8021-00130.\par 
This research has made use of the NASA/IPAC Extragalactic Database (NED), which is operated by the Jet Propulsion Laboratory, California Institute of Technology,
under contract with the National Aeronautics and Space Administration. This paper used the photoionization code Cloudy, which can be obtained from http://www.nublado.org. We thank Gary Ferland and associates, for the maintenance and development of Cloudy.

\section*{Data Availability}

Based on observations made with the NASA/ESA Hubble Space Telescope, and available from the Hubble Legacy Archive, which is a collaboration between the Space Telescope Science Institute (STScI/NASA), the Space Telescope European Coordinating Facility (ST-ECF/ESAC/ESA) and the Canadian Astronomy Data Centre (CADC/NRC/CSA).



\bibliographystyle{mnras}
\bibliography{HST_observation} 








\bsp	
\label{lastpage}
\end{document}